%% file: Main.tex
\documentclass[11pt, a4paper]{article}

\usepackage[english]{babel}
\usepackage{tikz}
\usepackage[compat=1.1.0]{tikz-feynman}
\usepackage{jheppub}
\usepackage{a4wide}
\usepackage{float}
\usepackage{bbold}
\usepackage{graphicx}
\usepackage[toc,page]{appendix}
\usepackage{amssymb}

\usepackage[output-decimal-marker={,},uncertainty-separator = {\pm}, separate-uncertainty = true]{siunitx}

\usepackage{rotating}
\usepackage{amsmath}
\usepackage{verbatim}
\usepackage{tabu}
\usepackage{rotating}
\usepackage{physics}
\usepackage{slashed}
\usepackage[bottom]{footmisc}
\usepackage{amsthm}
\usepackage{csquotes}
\usepackage{multirow}
\usepackage{url}
\usepackage{cleveref}
\usepackage{braket}
\usepackage{gensymb}
\usepackage{placeins}
\usepackage{makecell}
\usepackage{enumitem}
\usepackage[normalem]{ulem}
\newcommand{\ovbb}{$0\nu\beta\beta$\text{ }}

\newcommand{\lam}{\lambda}

\setlist{itemsep=2em}

\begin{document}

\title{Probing light sterile neutrinos in left-right symmetric models with displaced vertices and neutrinoless double beta decay}

\author[a,b]{Jordy de Vries,}\emailAdd{j.devries4@uva.nl}
\author[c]{Herbi K. Dreiner,}\emailAdd{dreiner@uni-bonn.de}
\author[a,b]{Jelle Groot,}\emailAdd{j.groot4@uva.nl}
\author[c]{Julian Y. Günther,}\emailAdd{guenther@physik.uni-bonn.de}
\author[d]{Zeren Simon Wang}\emailAdd{wzs@hfut.edu.cn}
\affiliation[a]{Institute for Theoretical Physics Amsterdam and Delta Institute for Theoretical
	Physics, University of Amsterdam, Science Park 904, 1098 XH Amsterdam, The
	Netherlands}
\affiliation[b]{Nikhef, Theory Group, Science Park 105, 1098 XG, Amsterdam, The Netherlands}
\affiliation[c]{Bethe Center for Theoretical Physics \& Physikalisches Institut der Universit\"{a}t Bonn, Nu{\ss}allee 12, 53115 Bonn, Germany}
\affiliation[d]{School of Physics, Hefei University of Technology, Hefei 230601, China}

\date{\today}

\vskip1mm
%%%%%%%%%%%%%%%%%%%%%%%%%%%%%%%%%%%%%%%%%%%%%%%%%%%%%%%%%%%%%%%%%%%%%%

\abstract{An investigation of relatively light (GeV-scale), long-lived right-handed neutrinos is performed within minimal left-right symmetric models using the neutrino-extended Standard Model Effective Field Theory framework.
Light sterile neutrinos can be produced through rare decays of kaons, $D$-mesons, and $B$-mesons at the Large Hadron Collider (\texttt{LHC}) and the Long-Baseline Neutrino Facility (\texttt{LBNF}) of Fermilab.
Their decays could result in displaced vertices, which can be reconstructed.
By performing Monte-Carlo simulations, we assess the sensitivities of the future \texttt{LHC} far-detector experiments \texttt{ANUBIS}, \texttt{CODEX-b}, \texttt{FACET}, \texttt{FASER(2)}, \texttt{MoEDAL-MAPP1(2)}, \texttt{MATHUSLA}, the recently approved beam-dump experiment \texttt{SHiP}, and the upcoming neutrino experiment \texttt{DUNE} at the \texttt{LBNF}, to the right-handed gauge-boson mass $M_{W_R}$ as functions of neutrino masses.
We find that \texttt{DUNE} and \texttt{SHiP} could be sensitive to right-handed gauge-boson masses up to $\sim 25$ TeV. We compare this reach to indirect searches such as neutrinoless double beta decay, finding that displaced-vertex searches are very competitive.}

\maketitle

\input{Sections/1_Introduction}

\input{Sections/2_mLRSM}

\input{Sections/3_nuSMEFT}

\input{Sections/4_SterileProduction}

\input{Sections/5_SterileDecay}

\input{Sections/6_Neutrinoless}

\input{Sections/7_ColliderAnalysis}

\input{Sections/8_NumericalResults}

\input{Sections/9_Conclusions}

\section*{Acknowledgment}

Financial support for H.K.D.~by the DFG (CRC 110, ``Symmetries and the Emergence of Structure in QCD'') is gratefully acknowledged.
JdV and JeG acknowledge support from the Dutch Research Council(NWO) in the form of a VIDI grant.

\bibliographystyle{JHEP}
\bibliography{references}

\end{document}

%% file: Sections/1_Introduction.tex
% !TEX root = ../Main.tex
\section{Introduction}

One of the big puzzles of the Standard Model (SM) is the mechanism behind the generation of neutrino masses. 
Astrophysical~\cite{Super-Kamiokande:1998kpq, SNO:2002tuh}  and laboratory~\cite{DoubleChooz:2011ymz, DayaBay:2012fng, RENO:2012mkc, T2K:2013ppw} neutrino flavor-oscillation experiments have unequivocally shown that SM neutrino masses are non-vanishing.
The SM electroweak symmetry group $SU(2)_L \otimes U(1)_Y$ cannot accommodate non-zero left-handed (LH) neutrino mass terms at the renormalizable level. 
However, adding right-handed (RH) neutrino fields, which do not transform under SM gauge symmetries, allows for non-zero renormalizable neutrino mass terms. 
Through a Yukawa term that couples a RH neutrino field to a SM Higgs field and a LH (SM) neutrino field, electroweak symmetry breaking (EWSB) can generate Dirac masses for the neutrinos in an analogous manner to the other (massive) SM fermions. 
However, a Majorana mass term is allowed for the RH neutrinos in this case.

The combination of both Dirac and Majorana mass terms leads to Majorana neutrino mass eigenstates and a split spectrum; this is usually referred to as the seesaw mechanism~\cite{Minkowski:1977sc, Yanagida:1979as, Gell-Mann:1979vob, Mohapatra:1979ia, Schechter:1980gr, Wyler:1982dd, Mohapatra:1986bd, Bernabeu:1987gr, Akhmedov:1995ip, Akhmedov:1995vm, Malinsky:2005bi}.
In addition to the three light active neutrinos, there must exist at least two (three if the lightest active neutrino is not massless) heavier Majorana neutrinos~\cite{Capozzi:2021fjo, Esteban:2020cvm, deSalas:2020pgw}. 
If these additional neutrinos have masses above the eV scale, their interactions are suppressed and are often denoted as sterile neutrinos or heavy neutral leptons (HNLs)~\cite{Shrock:1980vy,Shrock:1980ct,Shrock:1981wq} (for a review see Ref.~\cite{Abdullahi:2022jlv}).
Within this minimal scenario, the mass scale of sterile neutrinos is not well constrained. It can range from very heavy ($10^{15}$ GeV) to very light (eV), and the sterile-neutrino pairs can either be 
almost mass-degenerate or have a large mass hierarchy. 
A generic prediction, however, is that active neutrinos are Majorana and that processes violating lepton number ($L$) by two units are possible.

Recently, there has been much interest in sterile neutrinos with GeV-scale masses (see, for instance, Refs.~\cite{Atre:2009rg, Bondarenko:2018ptm, Helo:2013esa, Drewes:2019fou, Cottin:2022nwp, Liu:2023gpt}), especially in the context of searches for long-lived particles (LLPs) (see Refs.~\cite{Alimena:2019zri, Lee:2018pag, Curtin:2018mvb, Beacham:2019nyx} for recent reviews of LLPs).
In the minimal scenario where the only interactions between the sterile neutrinos and the SM particles are mediated by the active-sterile neutrino mixings $U^2$, the current bounds on the sterile neutrinos in this mass range hint that these sterile neutrinos, if existent, would be long-lived~\cite{Deppisch:2015qwa, Bryman:2019bjg, Abdullahi:2022jlv}.
Beam-dump, neutrino, and collider experiments produce mesons in very large numbers. These mesons can produce sterile neutrinos with masses in the MeV-GeV range through rare decays.
With a long lifetime, these sterile neutrinos travel a macroscopic distance before potentially decaying inside a detector volume placed a few to a few hundred meters away.
Past studies have investigated these minimal scenarios extensively and indicated that (proposed) future experiments such as \texttt{CODEX-b}~\cite{Helo:2018qej,Aielli:2019ivi}, \texttt{DarkQuest}~\cite{Batell:2020vqn}, \texttt{DUNE}~\cite{Krasnov:2019kdc, Ballett:2019bgd, Berryman:2019dme, Coloma:2020lgy, Moghaddam:2022tac, Plows:2022gxc, Gunther:2023vmz}, \texttt{ANUBIS}~\cite{Hirsch:2020klk, DeVries:2020jbs}, \texttt{SHiP}~\cite{SHiP:2015vad, Alekhin:2015byh, SHiP:2018xqw, DeVries:2020jbs}, and \texttt{MATHUSLA}~\cite{Helo:2018qej, Curtin:2018mvb, MATHUSLA:2020uve, DeVries:2020jbs}, could be sensitive to mixing angles $U^2$ close to or even reaching the type-I seesaw predictions $U^2 \sim m_\nu/m_N$, where $m_\nu$ denotes the sum of active-neutrino masses and $m_N$ the sterile neutrino mass, for $m_N$ in the GeV range.

More recently, studies on long-lived HNLs have been extended from minimal to non-minimal scenarios.
Sterile neutrinos might only appear sterile at relatively low energies because they interact through some decoupled beyond-the-SM (BSM) particles. 
Examples are models in which RH neutrinos interact with $Z'$ bosons~\cite{Deppisch:2019kvs, Chiang:2019ajm} or leptoquarks~\cite{Cottin:2021tfo, Bhaskar:2023xkm}. 
Effective-field-theory (EFT) techniques can efficiently describe such additional interactions.  
In particular, under the assumption that all BSM fields, except for the sterile neutrinos, are heavy compared to the electroweak scale, the neutrino-extended SM-EFT ($\nu$SMEFT)~\cite{delAguila:2008ir, Aparici:2009fh, Bhattacharya:2015vja, Liao:2016qyd} Lagrangian describes all interactions between sterile neutrinos and SM fields that are Lorentz- and gauge-invariant. 
The $\nu$SMEFT framework has been used to describe sterile-neutrino phenomenology at a wide range of experiments~\cite{DeVries:2020jbs, Cottin:2021lzz, Beltran:2021hpq, Zhou:2021ylt, Beltran:2022ast, Beltran:2023nli, Beltran:2023ksw, Duarte:2014zea, Duarte:2016caz, Han:2020pff, Barducci:2020ncz, Barducci:2020icf, Bischer:2019ttk, Li:2020lba, Li:2020wxi, Bolton:2020xsm, Dekens:2020ttz, Cirigliano:2021peb, Chala:2020pbn, Datta:2020ocb, Datta:2021akg}.

While efficient, a particular feature of the $\nu$SMEFT Lagrangian, which could be a disadvantage from certain perspectives, is that many possible operators exist even when considering operators only up to dimension six. 
Choosing specific benchmark scenarios where only a few operators are considered simultaneously has become customary. 
Typically, one operator is selected to allow for the production of a sterile neutrino, while another operator is chosen to induce the sterile neutrino to decay. 
While straightforward to implement, this procedure oversimplifies the possible underlying BSM scenarios and might not 
be realistic.
In particular, it allows for an effective decoupling between the production and decay of the sterile neutrinos, which is impossible in, for example, the minimal scenario, which we outline below.
Similarly, it is possible to consider sterile-neutrino interactions with only very specific quark and lepton flavors, which can altogether avoid the stringent limits from experiments probing lepton number violation such as neutrinoless double beta decay $(0\nu\beta\beta)$ searches~\cite{ CUORE:2019yfd, Majorana:2019nbd, CANDLES:2020wcr, GERDA:2019ivs, CUPID:2019gpc, Alenkov:2019jis, EXO-200:2019rkq,KamLAND-Zen:2022tow,KamLAND-Zen:2024eml}.
Again, this might not be very realistic.

To alleviate the above-mentioned problem of the $\nu$SMEFT, which is rooted in its intrinsic bottom-up approach, one can resort to UV-complete models of non-minimally coupled sterile neutrinos, taking a top-down perspective.
In this work, we investigate the minimal left-right symmetric model (mLRSM)~\cite{Mohapatra:1979ia, Pati:1974yy, Mohapatra:1974gc, Senjanovic:1975rk, Mohapatra:1980yp, Senjanovic:1978ev}.
In this model, RH neutrinos are charged under an $SU(2)_R$ gauge symmetry appended to the SM gauge symmetry and interact with so-called RH gauge bosons $W^\pm_R$ and $Z^\prime$.
The mLRSM explains the smallness of the non-vanishing active-neutrino masses with the seesaw mechanism~\cite{Minkowski:1977sc, Mohapatra:1979ia, Yanagida:1979as, Gell-Mann:1979vob, Glashow:1979nm, Mohapatra:1980yp}, provides a dark-matter candidate with the lightest sterile neutrino~\cite{Nemevsek:2012cd,Nemevsek:2023yjl}, and interprets the matter-antimatter asymmetry with leptogenesis~\cite{Fukugita:1986hr, Frere:2008ct, BhupalDev:2014hro, BhupalDev:2015khe, Dhuria:2015cfa, BhupalDev:2019ljp}.
Limits from collider searches for heavy gauge bosons and light sterile neutrinos indicate that $m_{W_R}\gtrsim 5\,$TeV~\cite{CMS:2021dzb, ATLAS:2023cjo} and $Z^\prime\gtrsim 4$ TeV~\cite{CMS:2023ooo}, such that the non-minimal interactions of RH neutrinos are feeble, and the nomenclature of sterile neutrinos still applies.
While RH neutrinos are often assumed to then also appear at the TeV scale, this is not necessarily true.
They could be much lighter, similar to how most SM fermions are light with respect to the electroweak gauge bosons (see Ref.~\cite{Mikulenko:2024hex} for a very recent study on the phenomenology of a light pseudo-Dirac pair of sterile neutrinos in the mLRSM). It has been argued that the relative lightness of RH neutrinos is required to avoid the strong CP problem \cite{Kuchimanchi:2014ota, Senjanovic:2020int, Li:2024sln}.
Furthermore, while the RH interactions of sterile neutrinos are feeble, they can be large compared to the minimal mixing (this is roughly the case if $m_{W_R} < m_W/\sqrt{U} \simeq m_W (m_N/m_\nu)^{1/4} \simeq 30\,$TeV, for $m_N = 1$\,GeV and $m_\nu=0.05$\,eV).

This work investigates the phenomenology of the mLRSM with GeV-scale sterile neutrinos.\footnote{Refs.~\cite{Li:2022cuq,Urquia-Calderon:2023dkf} studied the phenomenology of long-lived sterile neutrinos with masses of order $10$-$100$ GeV in the mLRSM at, respectively, the LHC and lepton colliders.}
We perform an explicit matching to the $\nu$SMEFT framework. We then follow the method of 
Refs.~\cite{DeVries:2020jbs, Gunther:2023vmz} to investigate the discovery potentials for the light 
long-lived sterile neutrinos with displaced-vertex (DV) searches to be performed at the \texttt{LHC} and 
the near detector of the \texttt{DUNE} experiment~\cite{DUNE:2020lwj, DUNE:2020ypp, DUNE:2020jqi, 
DUNE:2021cuw, DUNE:2021mtg} of the Long-Baseline Neutrino Facility (\texttt{LBNF}) at Fermilab.
For the \texttt{LHC},\!\footnote{We omit the \texttt{LHC} main experiments \texttt{ATLAS} and \texttt{CMS}, as their usual trigger requirement of a high-$p_T$ lepton or jet cannot be satisfied for the sterile neutrinos produced from meson decays as considered here. An exception, though, can arise at the \texttt{CMS} experiment where specific $B$-parking datasets~\cite{Bainbridge:2020pgi, CMS:2024syx,CMS:2024zhe} could be utilized~\cite{CMS:2024ita}, allowing for probing the sterile-neutrino scenarios studied in this work. Nevertheless, since the $B$-parking searches do not use the full datasets available, we choose not to include this possibility in the work.} we focus on the present and future experiments \texttt{ANUBIS}~\cite{Bauer:2019vqk}, \texttt{FASER} and \texttt{FASER2}~\cite{Feng:2017uoz,FASER:2018eoc}, \texttt{MATHUSLA}~\cite{Chou:2016lxi,Curtin:2018mvb,MATHUSLA:2020uve}, \texttt{FACET}~\cite{Cerci:2021nlb}, \texttt{CODEX-b}~\cite{Gligorov:2017nwh}, and \texttt{MoEDAL-MAPP1} and \texttt{MoEDAL-MAPP2}~\cite{Pinfold:2019nqj,Pinfold:2019zwp}, and the recently approved beam-dump experiment~\texttt{SHiP}~\cite{SHiP:2015vad,Alekhin:2015byh,SHiP:2018xqw,SHiP:2021nfo}. Since light sterile neutrinos with RH interactions can significantly impact $0\nu\beta\beta$ rates \cite{Bamert:1994qh, Blennow:2010th, Mitra:2011qr, deGouvea:2011zz, Li:2011ss, Barry:2013xxa, Ghosh:2013nya, Girardi:2013zra, Barea:2015zfa, Bolton:2019pcu, Asaka:2020wfo, Jha:2021oxl, Dekens:2024hlz}, we also make a detailed comparison to complementary probes provided by rare-meson-decay searches and $0\nu\beta\beta$ experiments.
Somewhat remarkably, we find that these very different probes have a comparable reach.

This paper is structured as follows.
In Sec.~\ref{sec:mLRSM}, we introduce the mLRSM.
In Sec.~\ref{sec:EFT}, we detail the $\nu$SMEFT framework. We discuss sterile-neutrino production and decay in the mLRSM in Secs.~\ref{sec:Nproduction} and~\ref{sec:Ndecay}, respectively.
We elaborate on the phenomenology of neutrinoless double beta decay in the mLRSM in Sec.~\ref{sec:0vbb}.
In Sec.~\ref{sec:simulations}, we explain the numerical simulation and computation procedures for collider analysis based on the theoretical scenarios discussed in the previous sections.
We present the numerical results in Sec.~\ref{sec:numericalResults} and provide our conclusions and an outlook for future work in Sec.~\ref{sec:conclusions}.

%% file: Sections/2_mLRSM.tex
% !TEX root = ../Main.tex
\section{Minimal left-right symmetric model}
\label{sec:mLRSM}
We provide a concise review of the mLRSM~\cite{Mohapatra:1979ia, Pati:1974yy, Mohapatra:1974gc, Senjanovic:1975rk, Mohapatra:1980yp, Senjanovic:1978ev} 
in this section.
The model extends the electroweak sector of the SM by introducing a gauge-group extension, placing RH fermions into doublets that transform under an $SU(2)_R$ gauge symmetry.
It augments the fermionic particle content of the SM by including RH neutrinos, which are charged under the $SU(2)_R$ gauge symmetry and couple to $SU(2)_R$ gauge bosons, known as "RH gauge bosons."
In combination with the $SU(3)_C$ gauge group for the strong interaction, the mLRSM gauge group is given by
\begin{align}
\label{eq:SymGroupmLRSM}
G_{\text{LR}} \,\equiv 
  \,SU(2)_L \otimes SU(2)_R \otimes U(1)_{B-L},
\end{align}
where ``LR'' stands for ``left-right'', with the associated fermionic particle content
\begin{equation}
\label{eq:currentEigenstates}
        (L_{L,R})_i=\begin{pmatrix} \nu_{L,R} \\ \ell_{L,R}   \end{pmatrix}_i, \quad   (Q_{L,R})_i=\begin{pmatrix} u_{L,R} \\  d_{L,R} \end{pmatrix}_i,
\end{equation}
where $i=1,2,3$ is a generation index, and the $L, R$ subscripts are associated with the left- and right-handed projection operators $P_{L, R}=(1\mp\gamma_5)/2$.
To ensure that the $G_{\text{LR}}$ symmetry group breaks down to $G_{\text{SM}}\equiv SU(2)_L \otimes U(1)_Y$, the scalar sector in the mLRSM should be an extension of the SM counterpart.
To break the $G_{\text{LR}}$ gauge symmetry, the following scalar triplets are introduced:
\begin{equation}
\label{eq:ScalarFields}
    \Delta_{L,R}\equiv\begin{pmatrix}
\delta_{L,R}^+ /\sqrt{2} & \delta_{L,R}^{++} \\
\delta_{L,R}^0 & -\delta_{L,R}^+ / \sqrt{2}
    \end{pmatrix},
\end{equation}
which transform as $\Delta_L \in (\textbf{3},\textbf{1},2)$ and $\Delta_R \in (\textbf{1},\textbf{3},2)$ under $G_{\text{LR}}$.
Furthermore, to govern the breaking of the electroweak symmetry, a scalar bi-doublet is introduced:
\begin{align}
     \quad \Phi \equiv \begin{pmatrix}
        \phi_1^0 & \phi_2^+ \\
        \phi_1^- & \phi_2^0
    \end{pmatrix},
\end{align} which transforms as $\Phi \in (\textbf{2},\textbf{2}^*,0)$.
The Yukawa interactions in the mLRSM are given by
\begin{equation}
\label{eq:yukawa}
    \mathcal{L}_Y=-\overline{Q}_L(\Gamma\Phi+{\tilde \Gamma}{}\tilde \Phi) Q_R - \overline{L}_L(\Gamma_l\Phi+{\tilde \Gamma}_l{\tilde \Phi})L_R-(\overline{L}_L^ci\sigma_2\Delta_LY_LL_L+\overline{L}_R^ci\sigma_2\Delta_RY_RL_R) + \text{h.c.}, 
\end{equation}
where ${\tilde \Phi}\equiv\sigma_2 \Phi^* \sigma_2$ with $\sigma_2$ the second Pauli matrix and $\Gamma,{\tilde \Gamma},\Gamma_l,{\tilde \Gamma}_l,Y_{L}$ and $Y_R$ are dimensionless Yukawa coupling matrices.
Furthermore, we have defined $\psi_{L,R}^c=P_{R,L}\psi^c$ with $\psi^c\equiv C\overline{\psi}^T$ with the charge conjugation matrix $C=-C^{-1}=-C^\dagger=-C^T$.  

We denote the most general form of the scalar fields' vacuum expectation values (vevs) that retain the $U(1)_{\rm EM}$ gauge invariance after spontaneous symmetry breaking as
\begin{equation}
\label{eq:ScalarFieldsVEV}
    \braket{\Phi}=\begin{pmatrix}
        \kappa / \sqrt{2} & 0 \\
        0 & \kappa^\prime e^{i\alpha} / \sqrt{2} 
    \end{pmatrix}  , \quad \braket{\Delta_L} = \begin{pmatrix}
        0 & 0 \\
        v_L e^{i\theta_L}/\sqrt{2} & 0
    \end{pmatrix}, \quad \braket{\Delta_R}=\begin{pmatrix}
        0 & 0 \\
        v_R /\sqrt{2} & 0
    \end{pmatrix},
\end{equation}
where all parameters  $\kappa,\,\kappa',\,v_L,\,v_R,\,\alpha,$ and $\theta_L$ are real-valued~\cite{Deshpande:1990ip}.
Imposing the relation $\sqrt{\kappa^2 + \kappa^{\prime 2}} = v$, with $v = 246$ GeV being the electroweak scale, ensures the correct masses for the SM gauge bosons $Z$ and $W^\pm \equiv W_L^\pm$. Consistency with electroweak precision observables \cite{Melfo:2011nx} and the absence of right-handed currents in experiments require $v_R\gg v\gg v_L$.
The phases $\alpha$ and $\theta_L$ induce CP violation, and, as elaborated upon below, the vevs $v_L$ and $v_R$ induce the Majorana neutrino masses.

After the EWSB, the leptonic Yukawa interactions induce the following neutrino-mass terms
\begin{align}
    \mathcal{L}_\nu = \frac{1}{2}\overline{N}M_\nu N^c + \text{h.c.}, \qquad M_\nu= \begin{pmatrix}
        M_L & M_D \\ 
        M_D^T & M_R
    \end{pmatrix},
\end{align}
where $M_\nu$ is a $6\times 6$ symmetric neutrino-mass  
matrix. We have defined $N$, which, with the generation
index, reads $N_i\equiv(\nu_{Li},\nu_{Ri}^c)^T$. For the Dirac and Majorana mass matrices, we note the relations
\begin{align}
  M_D \equiv (\kappa \Gamma_l+\kappa^\prime{\tilde \Gamma}_l e^{-i\alpha}) / \sqrt{2}, \quad 
  M_L \equiv \sqrt{2}Y_L^\dag v_L e^{-i\theta_L} , \quad 
  M_R \equiv\sqrt{2}Y_R v_R,
\end{align}
where, since $v_R\gg v\gg v_L$, one expects $M_R\gg M_{D}\gg M_L$, although this also depends on the choice of Yukawa couplings.

Throughout this work, it is most convenient to work in the neutrino mass eigenbasis instead of the flavor eigenbasis. For the rotation to the neutrino mass eigenbasis, we follow Ref.~\cite{Mitra:2011qr}. Diagonalization of the neutrino mass matrix is achieved through
\begin{align}
    U^\dagger M_\nu U^*= \text{diag}(m_\nu)\equiv (m_1,m_2,m_3,M_4,M_5,M_6) ,
\end{align}
where $U$ is a $6 \times 6$ unitary matrix. The mass eigenstates $\nu_{1,2,3}$ $(\nu_{4,5,6})$ for the active (sterile) neutrinos can be defined in terms of the current eigenstates as 
\begin{align}
    \nu = (\nu_1, \ldots, \nu_6)^T \equiv N_m + N^c_m=\nu^c,
\end{align} 
where we have defined $N\equiv U N_m$ with $N_m\equiv
(\nu_L^\prime, \nu_R^{\prime c})^T=U^\dagger (\nu_L, 
\nu_R^{c})^T$. 

We can express the mixing matrix $U$ as
\begin{equation}
\label{eq:U}
    U=\begin{pmatrix}
        U_{\text{PMNS}} & S \\
        T & U_R
    \end{pmatrix},
\end{equation}
where $U_{\text{PMNS}}$ is the usual $3 \times 3$ Pontecorvo-Maki-Nakagawa-Sakata (PMNS) LH neutrino-mixing matrix and $U_R$ its analog for RH neutrinos. The 
$3\times3$ matrices $S$ and $T$ dictate the mixing strengths between the left- and right-handed neutrino sectors, which are expressed with 
\begin{equation}
 S=RU_R , \quad T=-R^\dag U_{\text{PMNS}},
\end{equation}
where the $3\times3$ matrix $R\equiv M_DM_R^{-1}$ generates active-sterile neutrino mixing at the leading order.
Also at the leading order, the neutrino-mixing matrices $U_\text{\rm PMNS}$ and $U_R$ are defined through the relations
\begin{align}
\label{eq:UPMNSdef}
      U_{\rm PMNS}\,\widehat{m}_\nu\:U_{\rm PMNS}^T&= M_L - M_D M_R^{-1}M_D^T, \\
    \label{eq:URdef}
    U_R\widehat{M}_N\:U_R^T&=M_R,
\end{align}
where $\widehat{m}_\nu\equiv\text{diag}(m_1,m_2,m_3)$ and $\widehat{M}_N\equiv\text{diag}(M_4,M_5,M_6)$. For completeness, we note that the PMNS matrix at second order in active-sterile mixing gains non-unitary contributions through $(\mathbf{1}-\eta)U_{\rm PMNS}$, 
where $\eta\equiv RR^\dagger/2$. For our purposes, we can safely neglect these contributions since $\eta_{ij}<10^{-3}$ for every $i,j\in\{e,\mu,\tau\}$~\cite{Fernandez-Martinez:2016lgt}.  

Suppressing the generation indices, we write the left- and right-handed fields as 
\begin{align}
\label{eq:numassRot1}
    \nu_L &= (PU)P_L \nu = U_{\text{PMNS}} \nu_L^\prime + S \nu_R^{\prime c}, && \nu_L^c = (PU^*)P_R\nu = U_{\text{PMNS}}^* \nu_L^{\prime c} + S^* \nu_R^\prime \\
    \label{eq:numassRot2}
    \nu_R &= (P_sU^*)P_R\nu=T^*\nu_L^{\prime c}+U_R^*\nu_R^\prime , && \nu_R^c = (P_sU)P_L\nu = T \nu_L^\prime + U_R \nu^{\prime c}_R.
\end{align}
Note that, in comparison with the notations used in Refs.~\cite{Senjanovic:2011zz,Nemevsek:2012iq,Senjanovic:2016vxw}, we denote \(U_{\text{PMNS}} = V_L\) and \(U_R^* = V_R\).
In the equations above, the flavor-space projectors $P$ and $P_s$ are $6\times3$ matrices and defined as
\begin{equation}
    P \equiv \begin{pmatrix}
        \mathcal{I}_{3 \cross 3} & 0_{3 \cross 3}\end{pmatrix} , \quad P_s \equiv \begin{pmatrix} 0_{3 \cross 3} & \mathcal{I}_{3 \cross 3}
    \end{pmatrix}.
\end{equation}

A main characteristic of the mLRSM is the restoration of parity above $v_R$.
We must impose a discrete symmetry to have an LR symmetry in the Yukawa sector. There are two feasible choices~\cite{Maiezza:2010ic}: a generalized charge conjugation ($\mathcal{C}$) or a generalized parity ($\mathcal{P}$) scenario. The discrete field transformations are 
\begin{align}
    \label{eq:Cfieldtransform}
    &\mathcal{C}:\quad Q_L \leftrightarrow Q_R^c,\quad L_L \leftrightarrow L_R^c, \quad \Phi \leftrightarrow \Phi^T , \quad \Delta_L \leftrightarrow \Delta_R^*,\\
    \label{eq:Pfieldtransform}
    &\mathcal{P}:\quad Q_L \leftrightarrow Q_R,\quad L_L \leftrightarrow L_R, \quad \Phi \leftrightarrow \Phi^\dag , \quad \Delta_L \leftrightarrow \Delta_R\,.
\end{align}
% \HKD{where we have not considered the quark fields, as they are irrelevant to our analysis.}
Requiring the Lagrangian to be invariant under these two discrete symmetries respectively leads to the identities 
\begin{align}
    \label{eq:Csymmetry}
    &\mathcal{C}: \quad M_L=v_L/v_R M_Re^{-i\theta_L}, \quad M_D=M_D^T,\\
    \label{eq:Psymmetry}
    &\mathcal{P}: \quad M_L=v_L/v_R M_R^\dagger e^{-i\theta_L}, \quad M_D=M_D^\dagger,
\end{align}
reducing the number of free parameters.
The hermiticity of $M_D$ in the case of $\mathcal{P}$ is only valid in the limit of $(\kappa^\prime / \kappa) \sin \alpha \to 0$~\cite{Senjanovic:2015yea,Senjanovic:2014pva}.

The spontaneous breaking of the $SU(2)_{L}$ and $SU(2)_{R}$ 
leads to massive gauge bosons. 
For the charged gauge bosons, one expects the $SU(2)_R$ gauge bosons $W_R^\pm$ to mix with the SM gauge bosons which we label with $W_L^\pm$.
The mass eigenstates $W_{1,2}^\pm$ can be rotated into the flavor eigenstates $W_{L,R}^\pm$ through
\begin{equation}
\label{eq:WLRmasseigenstates}
    \begin{pmatrix}
        W_L^\pm \\
        W_R^\pm
    \end{pmatrix} = \begin{pmatrix}
        \cos \zeta & -e^{\mp i \alpha}\sin\zeta  \\
        e^{\pm i \alpha}\sin \zeta   & \cos \zeta
    \end{pmatrix} \begin{pmatrix}
        W_1^\pm \\
        W_2^\pm
    \end{pmatrix},
\end{equation}
where we have defined the mixing parameter \begin{align}
    \tan \zeta \equiv \frac{g_L \kappa\kappa^\prime }{g_R v_R^2}.
\end{align}
The physical charged gauge boson masses are
\begin{equation}
     M_{W_1}\simeq\frac{g_L v}{2}\simeq M_{W_L}, \quad  M_{W_2}\simeq\frac{g_R v_R}{\sqrt{2}}\simeq M_{W_R},
\end{equation}
where $g_{L,R}$ are the $SU(2)_{L,R}$ gauge couplings, respectively.
Once one imposes an LR symmetry, the gauge coupling constants $g_L, g_R$ and the SM $SU(2)_L$ gauge coupling $g$ become equal, $g=g_L=g_R$ which leads to $M_{W_L}/M_{W_R} \simeq v / (\sqrt{2}v_R) \ll 1$. The above relations result in 
\begin{align}
    \zeta\simeq\frac{\xi}{2(\xi^2+1)}\lambda, \quad \text{with }\lambda\equiv\left(\frac{M_{W_L}}{M_{W_R}}\right)^2, \quad \text{and }\xi\equiv\frac{\kappa^\prime}{\kappa}.
\end{align}

\subsection{Seesaw phenomenology}

We divide our discussion of the sterile-neutrino phenomenology into canonical type-I and type-II seesaw dominance scenarios.
The phenomenology further depends on the choice of the generalized LR symmetry -- either 
imposing the generalized parity $\mathcal{P}$ or the generalized charge conjugation $\mathcal{C}$.
We start with the more straightforward type-II seesaw scenario.

\subsubsection{Canonical type-II scenario}

Dirac Yukawa couplings are assumed to vanish in the canonical type-II seesaw dominance limit, hence $M_D = 0$.
This results in a block-diagonal neutrino mass matrix
\begin{align}
    M_N=\begin{pmatrix}
        M_L & 0 \\
        0 & M_R 
    \end{pmatrix}.
\end{align} 
The physical neutrino masses $\widehat{m}_\nu$ and $\widehat{M}_N$ can be obtained by diagonalizing the Majorana mass matrices $M_L$ and $M_R$. From Eqs.~\eqref{eq:UPMNSdef} and \eqref{eq:URdef}, we find
\begin{align}
    U_{\text{PMNS}}^\dagger M_L U_{\text{PMNS}}^*=\widehat{m}_\nu, \qquad U_R^\dagger M_R U_R^*=\widehat{M}_N.
\end{align}
Employing the relations between $M_L$ and $M_R$ given in Eqs.~\eqref{eq:Csymmetry} and \eqref{eq:Psymmetry}, it is 
straightforward to obtain the relations between the RH neutrino mixing matrix $U_R$ and the PMNS matrix for the two choices of discrete LR symmetries:
\begin{align}
\label{eq:CPtypeII}
\mathcal{C}: U_R =U_\text{PMNS}e^{i\theta_L/2},\qquad
    \mathcal{P}: U_R =U_\text{PMNS}^* e^{-i\theta_L/2}.
\end{align}
Both of these cases lead to the neutrino mass relation 
\begin{align}
    \label{eq:MNtom}
    \widehat{M_N}=\frac{v_R}{v_L}\widehat{m}_\nu.
\end{align}

In the Normal Hierarchy (NH) scenario, we have the active-neutrino mass ordering $m_1<m_2<m_3$, with
\begin{align}
\label{eq:activeNHrelations}
    m_{2,3}=(m_1^2+\Delta m_{21,31}^2)^{1/2}, 
\end{align}
where the squared mass differences $\Delta m_{ij}^2\equiv m_{i}^2-m_{j}^2$ have been experimentally determined and are tabulated in Tab.~\ref{tab:activeneutrinolimits}.
On the other hand, the Inverted Hierarchy (IH) scenario has $m_3 < m_1 < m_2$, with
\begin{align}
    m_{1}=(m_3^2+\Delta m_{31}^2)^{1/2}, \quad m_{2}=(m_3^2+\Delta m_{21}^2-\Delta m_{31}^2)^{1/2}.
\end{align}

\begin{figure}[t!]
    \centering
    \includegraphics[width=\textwidth]{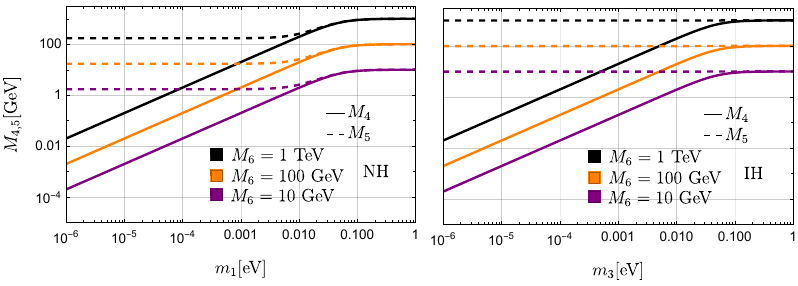}
    \vspace*{-0.6cm}
    \caption{Sterile-neutrino masses $M_4$ and $M_5$ as a function of the lightest neutrino mass in both the NH (left) and the IH (right) for different choices of heaviest sterile neutrino mass $M_6$.}
    \label{fig:m45vsm1plot}
\end{figure}

Explicitly, Eq.~\eqref{eq:MNtom} implies the relations
\begin{align}
\label{eq:m1m4massRelations}
    \text{NH}: M_{4,5}=\frac{m_{1,2}}{m_3}M_6, \qquad \text{IH}: M_{4,5}=\frac{m_{3,1}}{m_2}M_6\,,
\end{align}
between the active- $(m_{1,2,3})$ and sterile-neutrino mass eigenstates $(M_{4,5,6})$, irrespective of the choice of 
the additional discrete LR symmetry. The kinematic 
accessibility of the two lightest sterile neutrinos,
$\nu_{4,5}$ with masses $M_{4,5}$, depends on the choice of hierarchy and the choice of $m_1$ and $M_6$. The masses of the sterile neutrinos can be expressed in terms of the active-neutrino masses once we fix the highest sterile-neutrino mass and decide the choice of neutrino-mass hierarchy. The relations in Eq.~\eqref{eq:m1m4massRelations} are depicted in Fig.~\ref{fig:m45vsm1plot}. 

Neutrino-oscillation experiments have measured active-neutrino mixing angles and squared mass differences between the active-neutrino mass eigenstates.
We parameterize the PMNS matrix as~\cite{Workman:2022ynf}
\begin{align}
    U_{\rm PMNS}=\begin{pmatrix}
       1 & 0 & 0 \\
       0 & c_{23} & s_{23} \\
       0 & -s_{23} & c_{23}
    \end{pmatrix} \begin{pmatrix}
       c_{13} & 0 & s_{13}e^{-i\delta} \\
       0 & 1 & 0 \\
        -s_{13}e^{i\delta} & 0 & c_{13}
    \end{pmatrix}\begin{pmatrix}
       c_{12} & s_{12} & 0\\
       -s_{12} & c_{12} & 0\\
       0 & 0 & 1
    \end{pmatrix}\begin{pmatrix}
        1 & 0 & 0 \\
        0 & e^{i\lambda_1} & 0 \\ 
        0 & 0 & e^{i\lambda_2}
    \end{pmatrix},
\end{align}
where $\lambda_{1}$ and $\lambda_{2}$ are unknown Majorana phases.
We tabulate the current experimental values of these PMNS parameters in Tab.~\ref{tab:activeneutrinolimits}.
\setlength\extrarowheight{5pt}

\begin{table}[t!]
\resizebox{\textwidth}{!}{
\begin{tabular}{l|llllll}
 & $\Delta m_{21}^2 [10^{-5} \text{ eV}^2]$ & $|\Delta m_{31}^2| [10^{-3} \text{ eV}^2]$ & $\sin^2\theta_{12}$ & $\sin^2\theta_{23}$  & $\sin^2\theta_{13}$ & $\delta [\degree]$ \\
 \hline \hline
NH & $7.42_{-0.20}^{+0.21}$ & $2.514_{-0.028}^{+0.027}$ & $0.304^{+0.013}_{-0.012}$ & $0.570^{+0.018}_{-0.024}$ & $0.02221^{+0.00068}_{-0.00062}$ & $195^{+51}_{-25}$ \\
IH & $7.42_{-0.20}^{+0.21}$ & $2.497_{-0.028}^{+0.028}$ & $0.304^{+0.013}_{-0.012}$ & $0.575^{+0.017}_{-0.021}$ & $0.02240^{+0.00062}_{-0.00062}$ & $286^{+27}_{-32}$
\end{tabular}}
\caption{Active-neutrino oscillation parameters with squared mass differences $\Delta m_{ij}^2\equiv m_i^2-m_j^2$ and $\Delta m_{31}^2>0$ $(< 0)$ in the NH (IH)~\cite{Esteban:2020cvm}.}
\label{tab:activeneutrinolimits}
\end{table}

The decay rates of sterile neutrinos into SM particles -- and vice versa -- will depend on the sterile-neutrino masses, RH gauge-boson masses, and mixing angles in the diagonalization matrix $U$; see Eq.~\eqref{eq:U}.
In the generalized parity symmetry $\mathcal{P}$ scenario, a lower limit on the RH gauge-boson masses can be determined through low-energy experimental bounds, independent of the sterile-neutrino masses. In principle, measurements of CP-violation in kaon decays strongly limit $M_{W_R}\geq 17$ TeV~\cite{Bertolini:2019out, Dekens:2021bro}. The bound weakens to $M_{W_R}\geq5.5$ TeV when applying an additional Peccei-Quinn mechanism since additional freedom appears in such a non-trivial model extension. This weaker bound is comparable to limits from direct searches at the \texttt{LHC}~\cite{Dekens:2021bro}.

\subsubsection{Canonical type-I Scenario}

We continue by exploring the mLRSM in the type-I seesaw scenario. In this case, a non-zero Dirac mass matrix $M_D$ is allowed, and the Majorana mass matrix $M_L$ is zero by assuming $v_L=0$. Unlike the type-II seesaw scenario, sterile-neutrino decay and production rates now gain additional contributions through minimal mixing between the active and sterile neutrinos. 

A completely general study of the type-I seesaw scenario is beyond the scope of this work. Instead, we follow Ref.~\cite{deVries:2022nyh} and assume an explicit form of the RH neutrino mixing matrix
\begin{align}
\label{eq:LRtype1}
    \mathcal{C}: \quad U_R=U_{\rm PMNS}, \qquad \mathcal{P}: \quad U_R=U_{\rm PMNS}^*\,.
\end{align}
 In the $v_L=0$ limit, this leads to the following expressions for the Dirac mass matrix: 
\begin{align}
        \mathcal{C}: \quad M_D&=U_{\rm PMNS}\widehat{M}_N\sqrt{-\widehat{M}_N^{-1}\widehat{m}_\nu }U^T_{\rm PMNS},\\
    \mathcal{P}: \quad M_D&=U_{\rm PMNS}\widehat{M}_N\sqrt{-\widehat{M}_N^{-1}\widehat{m}_\nu }U^\dagger_{\rm PMNS}.
\end{align}
As was the case in the type-II seesaw model, in the limit of $\xi\sin\alpha\to0$~\cite{Senjanovic:2014pva, Senjanovic:2015yea}, we obtain the Dirac mass matrix equality in the case of a generalized $\mathcal{P}$ symmetry.
To leading order, we can express the active-sterile neutrino mixing matrix $R=M_DM_R^{-1}$ as
\begin{align} 
\label{eq:Rtype1}
    \mathcal{C}:\quad R=iU_{\rm PMNS}\widehat{R}U_{\rm PMNS}^\dagger,\qquad
    \mathcal{P}:\quad R=iU_{\rm PMNS}\widehat{R}U_{\rm PMNS}^T,
\end{align} where we have defined
\begin{align}
    \widehat{R}\equiv\text{diag}\left(\sqrt{\frac{m_1}{M_4}},\sqrt{\frac{m_2}{M_5}},\sqrt{\frac{m_3}{M_6}}\right).
\end{align}
Now, we can obtain our final expression for the neutrino mixing matrix $U$ in the type-I Seesaw scenario by substituting Eqs.~\eqref{eq:LRtype1} and~\eqref{eq:Rtype1} into Eq.~\eqref{eq:U}.
For the two choices of LR symmetry, we have
\begin{align}
\label{eq:CtypeI}
    \mathcal{C}:\quad U&=\begin{pmatrix}
        U_{\rm PMNS} & iU_{\rm PMNS} \widehat{R} \\
        iU_{\rm PMNS}\widehat{R} & U_{\rm PMNS}
    \end{pmatrix}, \\
    \label{eq:PtypeI}
    \mathcal{P}:\quad U&=\begin{pmatrix}
        U_{\rm PMNS} & iU_{\rm PMNS}\widehat{R} \\
        iU^*_{\rm PMNS} \widehat{R} & U^*_{\rm PMNS}
    \end{pmatrix},
\end{align}
which are denoted as $\mathcal{C}$- and $\mathcal{P}$-symmetric, respectively.

It is important to emphasize that, unlike in the type-II seesaw phenomenology, there are no direct relations between the active- and sterile-neutrino masses.
Hence, the sterile-neutrino masses $M_{4,5,6}$ are free parameters in the type-I seesaw scenario and could thus vary over a broad mass range.
In principle, $M_{4,5,6}$ could all be kinematically allowed in searches for long-lived HNLs.
We provide an overview of independent parameters given the seesaw scenarios under consideration in Tab.~\ref{tab:mLRSMparameters}.

\begin{table}[t!]
\begin{tabular}{l|lll}
                  & Seesaw model & RH neutrino mixing matrix & Relevant parameters \\
                  \hline \hline
\multirow{2}{*}{$\mathcal{P}$} & Type-I  & $U_R=U_{\text{PMNS}}^*$ & $M_{W_R},\xi,\alpha,\lambda_1,\lambda_2,\theta_L,m_1,M_4,M_5,M_6$ \\
                               & Type-II & $U_R=U_{\text{PMNS}}^*e^{-i\theta_L/2}$ & $M_{W_R},\xi,\alpha,\lambda_1,\lambda_2,\theta_L,m_1,M_6$ \\
\multirow{2}{*}{$\mathcal{C}$} & Type-I &  $U_R=U_{\text{PMNS}}$   &        $M_{W_R},\xi,\alpha,\lambda_1,\lambda_2,\theta_L,m_1,M_4,M_5,M_6$           \\
                  &      Type-II   &   $U_R=U_{\text{PMNS}}e^{i\theta_L/2}$    &     $M_{W_R},\xi,\alpha,\lambda_1,\lambda_2,\theta_L,m_1,M_6$ 
\end{tabular}
\caption{Summary of independent parameters in the mLRSM scenarios under consideration in the NH case. The replacement $m_1\to m_3$ gives the corresponding summary for the IH case.}
\label{tab:mLRSMparameters}
\end{table}

%% file: Sections/3_nuSMEFT.tex
% !TEX root = ../Main.tex
\section{Standard model effective field theory extended by sterile neutrinos}
\label{sec:EFT}

\subsection{The effective neutrino Lagrangian}

In the mLRSM, the masses of the RH scalars and gauge bosons depend on the $SU(2)_R$ vev $v_R$.
However, the sterile neutrinos have masses $M_i$, with $ i=4,5,6$, which we assume to be significantly smaller. Since $M_i\ll v_R$, there is a separation of scales, which suggests applying EFT techniques.
At the scale $v_R$, we integrate out heavy mLRSM degrees of freedom and match to the $\nu$SMEFT Lagrangian~\cite{Liao:2016qyd}.
The resulting EFT contains singlet $\nu_R$ fields in addition to SM fields and accommodates all effective operators explicitly invariant under the SM gauge group.  

Since we assume that $v_R\gg v_L$, additional (charged) Higgs scalars are too heavy to contribute to sterile-neutrino production and decay rates significantly.
The relevant terms in the charged-current dim-6 $\nu$SMEFT Lagrangian containing at least one sterile neutrino are 
\begin{align}
\begin{split}
\label{eq:L6CCnuSMEFT}
\mathcal{L}^{(6){\rm CC}}_{\nu {\rm SMEFT}}=\sum_{i}\Bigg\{ 
C^{(6)}_{\varphi \nu l} \left[(D_\mu \varphi)^\dagger i {\tilde \varphi} \right]  \left[\overline{l}_R^i\gamma^\mu \nu_R^i\right] +\sum_{j,k}\left( C^{(6)}_{du\nu l,jk} \left[ \overline{u}_R^j\gamma^\mu d_R^k \right]\left[\overline{l}_R^i\gamma_\mu \nu_R^i \right] \right) \Bigg\}+ {\rm h.c.},
\end{split}
\end{align}
where $\nu_{R}$ and $l_R$ are right-chiral neutrino and lepton fields and ${u}_{R}$ and $d_R$ are right-chiral up- and down-type quark fields.
Furthermore, $i,j,k=\{1,2,3\}$ are flavor-generation indices and $C_{du \nu l}^{(6)},C_{\varphi \nu l}^{(6)}$ are $3\times3$ matrices in lepton-flavor space that are given by
\begin{align}
\label{eq:C6CCnuSMEFT}
   C^{(6)}_{du\nu l,jk}=-\frac{V^R_{jk}}{v_R^2},\qquad  C^{(6)}_{\varphi \nu l}= -\frac{1}{v_R^2 }\frac{2\xi}{1+\xi^2} e^{-i\alpha},
\end{align}
with the ratio of vevs $\xi\equiv \kappa^\prime/\kappa$ (cf.~Eq.~\eqref{eq:ScalarFieldsVEV}) and $V_{ij}^R$ are elements of the RH $3\times3$ quark-mixing matrix.

In $\mathcal{P}$-symmetric mLRSMs (cf.~Eq.~\eqref{eq:PtypeI}), 
these are equal to elements of the LH (CKM) quark mixing matrix in the limit of $\xi\sin\alpha \to 0$.
In $\mathcal{C}$-symmetric mLRSMs (cf.~Eq.~\eqref{eq:CtypeI}),  $V^R$ equals the complex conjugated CKM matrix $V^*$, up to additional phases.
In this work, the amplitudes for meson decays are governed by, at most, one diagram containing one quark transition.
Hence, only the norms of the quark mixing matrix elements are relevant, which are equivalent in both the $\mathcal{C}$- and $\mathcal{P}$-symmetric mLRSMs.

After we apply the matching procedure to the $\nu$SMEFT Lagrangian, we evolve the higher-dimensional operators to the EW scale $v$, where we integrate out heavy SM fields (top quark, Higgs boson, and $W^\pm$- and $Z$-bosons). 
We match the resulting terms at the tree level to a $SU(3)_C \otimes U(1)_{\rm EM}$-invariant neutrino-extended low-energy EFT ($\nu$LEFT). The vector-current operators in Eq.~\eqref{eq:L6CCnuSMEFT} have vanishing QCD anomalous dimensions at one loop and hence do not evolve under QCD between $v_R$ and $v$ up to negligible electroweak corrections.

The relevant charged-current operators are 
\begin{align}
\begin{split}
\label{eq:LEFTCC}
\mathcal{L}_{\nu \text{LEFT}}^{(6){\rm CC}}=&\frac{2G_F}{\sqrt{2}}\sum_{i,j,l}\Bigg\{\sum_{k}\Big(\overline{u}_L^i\gamma^\mu d_L^j\left[C^{(6)}_{\text{VLL1},ijkl}\overline{e}_L^k \gamma_\mu\nu^l+C^{(6)}_{\text{VLR1},ijkl}\overline{e}_R^k\gamma_\mu \nu^l\right]\\&+ C^{(6)}_{\text{VRR},ijkl} \overline{u}^i_R\gamma^\mu d_R^j\overline{e}_R^k\gamma_\mu \nu^l \Big) +\overline{\nu}_L^i\gamma^\mu e^i_L\left[C^{(6)}_{\text{VLL2},jl}\overline{e}_L^j\gamma_\mu\nu^l+C^{(6)}_{\text{VLR2},jl}\overline{e}_R^j\gamma_\mu\nu^l \right]\Bigg\} ,
\end{split}
\end{align}
where we have rotated to the neutrino mass basis, cf.~Eqs.~\eqref{eq:numassRot1} and~\eqref{eq:numassRot2}. 
In the equation above, $G_F=(\sqrt{2}v^2)^{-1}$ is the Fermi constant, ${u}_{L,R}$ and ${d}_{L,R}$ are chiral up- and down-type quark fields, $e_{L,R}$ are charged lepton fields, and $i,j,k=\{1,2,3\}$ are flavor indices, whereas $l=\{1,\dots,6\}$ denotes the neutrino mass eigenstates.
The Wilson coefficients are given by
\begin{align}
    {C_{\text{VLL1},ijkl}^{(6)}}  &= -2 V_{ij}(PU)_{kl}, && {C_{\text{VLL2},kl}^{(6)}}  = \frac{C_{\text{VLL1},ijkl}^{(6)}}{V_{ij}},
    \label{eq:CVLL12}\\
    {C_{\text{VLR1},ijkl}^{(6)}}  &=V_{ij} v^2 (P_s U^*)_{kl}  \: C^{(6)}_{\varphi \nu l},
    \label{eq:CVLR12}
    && {C_{\text{VLR2},kl}^{(6)}}  =  \frac{C_{\text{VLR1},ijkl}^{(6)}}{V_{ij}},\\
    \label{eq:CVRR}
    {C_{\text{VRR},ijkl}^{(6)}}  &=   v^2 (P_s U^*)_{kl} \: C_{du\nu l,ij}^{(6)}.
\end{align}

The relevant neutral-current interactions are 
\begin{align}
    \mathcal{L}^{(6){\rm NC}}_{\nu {\rm LEFT}} = -\frac{4G_F}{\sqrt{2}} \sum_{i,j,k} \eta_{Z}^{jk}(f) C^{(6)}_{Z,ijk}({f})\left[\overline{\nu}_L^j\gamma_\mu \nu^i\right] \left[\overline{{f}}^k\gamma^\mu {f}^k\right],
\end{align}
where $i=\{1\dots6\}$ are flavor indices for the neutrino mass eigenstates, $j,k=\{1,2,3\}$ are flavor indices for fermion fields $f\in \{u_L,d_L,u_R,d_R,l_L,\nu_L,l_R\}$, and 
\begin{align}
    C^{(6)}_{Z,ij}({f})&=(P U)_{ji}\left[T^3_L-\sin^2 \theta_W Q\right],
\end{align}
with electric charge $Q$ and the isospin projection $T^3_L=\tau^3/2$ for LH doublets.
For the symmetry factor, we have  $\eta_{Z}^{ij}({\nu_L}) =\frac{1}{2}(2-\delta_{ij})$ and $\eta_{Z}^{ij}({f}) =1$ otherwise.

Since the vector-current operators in Eq.~\eqref{eq:LEFTCC} have vanishing QCD anomalous dimensions, they do not evolve between $v$ and sterile-neutrino masses $M_i$ when $v>M_i>\lambda_\chi\sim \text{GeV}$.
The dim-6 neutral-current and charged-current $\nu$LEFT Lagrangians are sufficient to calculate all sterile-neutrino production and decay rates in this work.

All the terms in the dim-6 $\nu$LEFT Lagrangian are lepton-number conserving.
Nevertheless, they allow for non-zero contributions to $0\nu\beta\beta$ processes, which induce lepton-number violation through Majorana masses $M_i$ originating from virtual neutrino propagators.
We further elaborate on this point in Sec.~\ref{sec:0vbb}. 

%% file: Sections/4_SterileProduction.tex
% !TEX root = ../Main.tex
\section{Sterile-neutrino production}
\label{sec:Nproduction}

In this section, we investigate the production rates of sterile neutrinos.
We focus on the production via decays of mesons, which are copiously produced at the interaction points (IPs) of the experiments considered in this work.
The production of sterile neutrinos with mass $\mathcal{O}(\text{GeV})$ via the decays of Higgs, $W$- and $Z$-boson, if existent, is subdominant as a result of the relatively smaller production cross sections of these bosons, and is hence not considered here.
This analysis studies leptonic and semi-leptonic decays of $K^\pm$, $K_L$, $K_S$, $D^\pm, D^0, D_s, B^\pm, B^0$, and $B_s$ mesons.
We neglect subdominant contributions from $B_c$, $J/\Psi$, and $\Upsilon$ mesons, though they might extend the upper mass reach of the sterile neutrinos to a minor extent. 

\subsection{Minimal mixing scenario}

We start by analyzing the phenomenology of a scenario where sterile neutrinos are only produced through minimal mixing with the SM neutrinos.
In this scenario, there are no RH gauge-boson contributions, and the production of the sterile neutrinos occurs exclusively via LH currents.
The only relevant Wilson coefficients are ${C_{\text{VLL1},ijkl}^{(6)}}  = -2 V_{ij}(PU)_{kl}$, where $V_{ij}$ are CKM matrix elements and $U$ the neutrino mixing matrix.

It is simplest to assume that only the lightest sterile neutrino is kinematically available to be produced from the meson decays. From this point forward, for notational clarity, we will denote the sterile neutrino mass eigenstates as $N_4, N_5$ and $N_6$.
The sterile-neutrino mass range we consider is $0< M_4 < m_{B_s} \sim 5.37 \text{ GeV}$.
Unless explicitly stated otherwise, we follow Ref.~\cite{deVries:2022nyh} for our decay-rate calculations, including the phase-space integration, the relevant form-factor parameterizations, and the decay constants.

We consider the branching ratios of mesons decaying into final states that include exactly one sterile neutrino.
This is reasonable in the minimal mixing scenario owing to the smallness of the relevant neutrino mixing angles.
Branching ratios in this scenario have been calculated and discussed extensively in the literature; see e.g.~Refs.~\cite{Bondarenko:2018ptm, deVries:2022nyh}.
In the considered sterile-neutrino production processes, either no meson, a pseudoscalar meson, or a vector meson accompanies the sterile neutrino in the final state.

\begin{figure}[t]
    \centering
    \includegraphics[width=\textwidth]{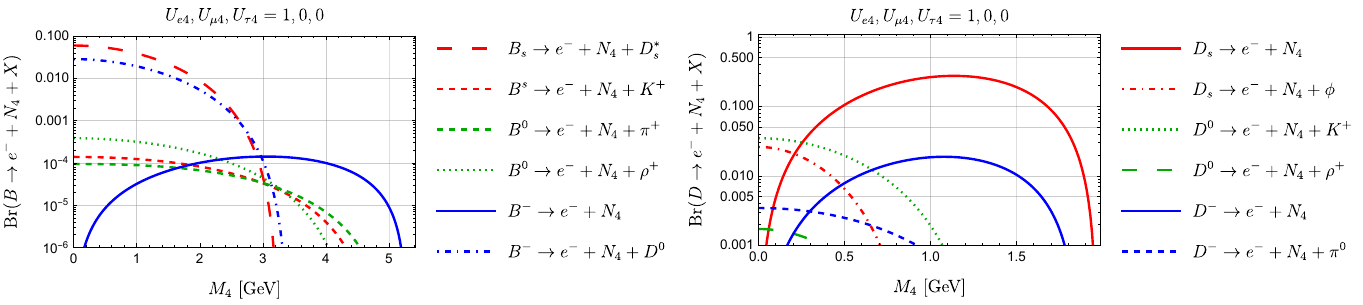}
        \vspace*{-0.6cm}
    \caption{A selection of branching ratios for decays of $B$ and $D$ mesons into a sterile neutrino in the minimal scenario.
    }
    \label{fig:DmesonSM}
\end{figure}

In Fig.~\ref{fig:DmesonSM}, the branching ratios of $B, D \to e^- + N_4 + X$ processes are shown for mixing angles $(U_{e4},U_{\mu 4},U_{\tau 4})=(1,0,0)$, where $X$ denotes the potential final-state meson.
We have arbitrarily set the mixing angle $U_{e4}=1$ since the decay rates in all depicted processes scale with an overall factor $|U_{e4}|^2$.
From Fig.~\ref{fig:DmesonSM}, it is evident that decays with no mesons, pseudoscalar mesons, or vector mesons in the final-state configuration all give dominant contributions at different points in the sterile-neutrino mass range.
Hence, we take all these decays into account in our further analysis.
Our results agree with those derived in previous works~\cite{Bondarenko:2018ptm, DeVries:2020jbs}.

\subsection{Left-right symmetric models}

We continue by analyzing sterile-neutrino production in the mLRSM scenarios.
We first assume a type-II seesaw scenario, where Dirac masses $M_D$ are set to zero, and there is no minimal mixing between the left- and right-handed neutrino sectors.
Hence, all active-sterile neutrino mixing angles are zero, and there are no purely LH current interactions with sterile neutrinos since $C_{\rm VLL1}^{(6)}=0$.
Neutral-current sterile-neutrino production via semi-leptonic meson decays is loop-suppressed because the $Z$ and $Z^\prime$ boson interactions are quark-flavor diagonal.
An example of such a decay would be $B_s \to K_S/\phi + N_{4,5} + N_{4,5}$.
Because of this loop suppression, such decay channels are neglected.
This implies that the right-handed charged gauge boson $W_R$ is the only BSM gauge boson providing non-negligible contributions to the meson decay rates.

Since we neglect neutral-current processes, the sterile-neutrino production rates are similar for all sterile-neutrino mass eigenstates with identical masses.
That is, in this limit, the only difference between the production rates of $N_4, N_5$, and $N_6$ via meson decays arises from an overall factor $|U_R|_{kl}^2$ containing the appropriate RH neutrino 
mixing-matrix angle.
This implies
\begin{align}
    \frac{\text{Br}(M\to \ell^- + N_{k_1} + X)}{\text{Br}(M\to \ell^- + N_{k_2} + X)}=\frac{\abs{U_{R}}^2_{k_1l}}{\abs{U_{R}}^2_{k_2l}},
\end{align}
where the meson $M\in\{B,D,K\}$ and the charged lepton $\ell\in\{e,\mu,\tau\}$. 
Since only the absolute squares of the mixing angles are relevant, there is no dependence on the complex phases $\lam_1,\lam_2$, and $\theta_L$ that appear in $U_R$.
The choice of the discrete LR symmetry is also irrelevant to the meson decay rates in this scenario.
This greatly simplifies the available parameter space.
The choice of the heaviest neutrino mass, $M_6$, fixes the mass relation between the active-neutrino mass $m_1$ and the lightest sterile-neutrino mass $M_4$ and does not impact the meson decay rates as long as $M_6$ is large enough to make $N_5$ kinematically unavailable, cf.~Eq.~\eqref{eq:m1m4massRelations}.
For simplicity, we set the CP-violating phase $\alpha=0$. From Tab.~\ref{tab:mLRSMparameters}, in the limit where only $N_4$ is kinematically available, it is then apparent that only two relevant benchmark parameters remain. 
These are the $\xi=\kappa^\prime/\kappa$, which determines the $W_L$-$W_R$ 
mixing strength, and $M_{W_R}$, which is the mass of the RH 
charged gauge boson. 

\begin{figure}[t!]
    \centering
    \includegraphics[width=\textwidth]{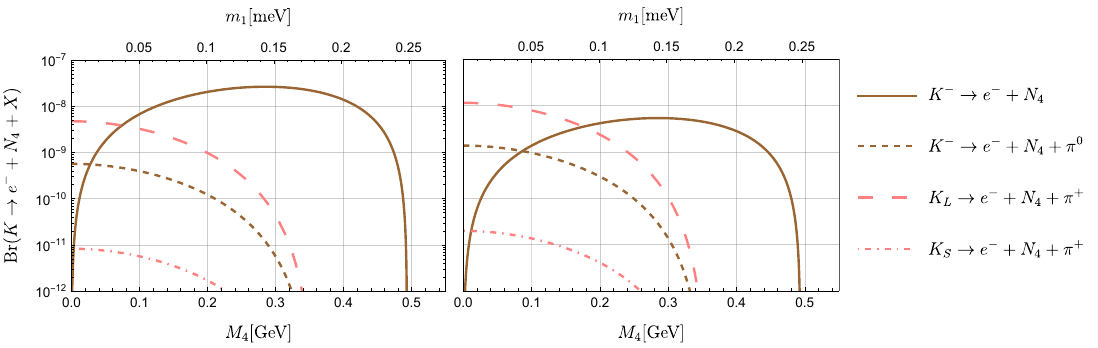}
    \includegraphics[width=\textwidth]{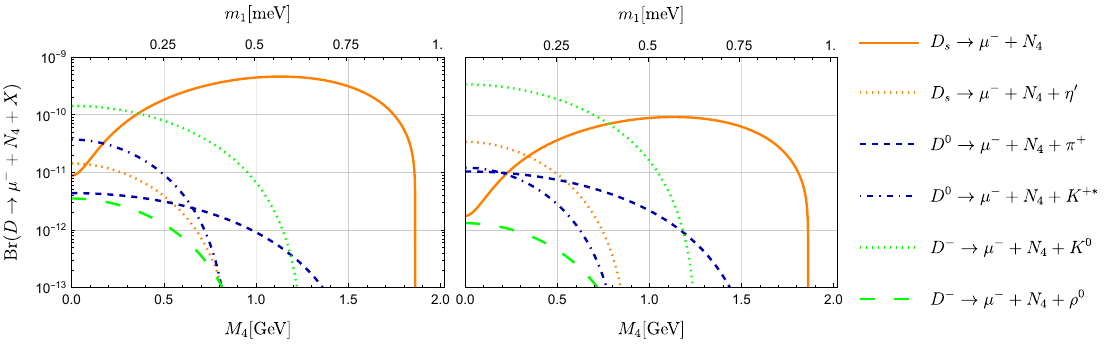}
    \includegraphics[width=\textwidth]{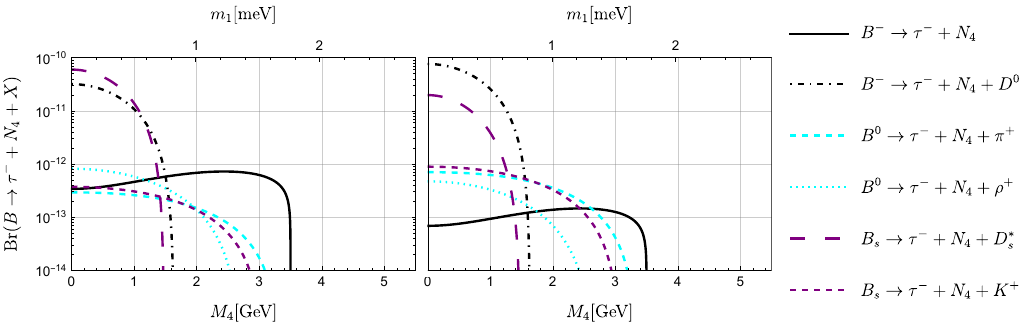}
    \vspace*{-0.6cm}
    \caption{A selection of branching ratios for $M\to l^- + N_4 + X$ decay processes, where $M=\{B, D, K\}$ and $l=\{e,\mu,\tau\}$, in mLRSM scenarios for $M_6=100$ GeV and $M_{W_R}=7$ TeV. We work in the NH and have set the $W_L$-$W_R$ mixing parameter $\xi=0$ ($\xi=0.3$) in the left (right) panels.}
    \label{fig:BrBtauNLRplot}
\end{figure}

In Fig.~\ref{fig:BrBtauNLRplot}, representative branching ratios are shown for $M\to \ell^- + N_4 + X$ decay processes.
We have fixed the RH gauge-boson mass at a relatively small value of $M_{W_R}=7$ TeV and considered both cases of $\xi=0$ and $\xi=0.3$ for the $W_L$-$W_R$ mixing parameter.
In this scenario, all decay rates that produce sterile neutrinos scale with $M_{W_R}^{-4}$.
The heaviest sterile-neutrino mass has been set to $M_6=100$\,GeV, so the sterile-neutrino mass eigenstate $M_5$ is kinematically unavailable.
Choosing a larger value for $M_6$ would not impact the dependence of the meson decay rates on $M_4$, but it would shift the $m_1$ axes through Eq.~\eqref{eq:m1m4massRelations}.
Leptonic and semi-leptonic meson decays into final-state pseudoscalar or vector mesons dominate the sterile-neutrino production rates at different $M_4$ and are hence all considered in the further analysis. 

Furthermore, a non-zero value of $\xi$ amplifies the pseudoscalar-meson decay rates into a sterile neutrino, a charged lepton, and a pseudoscalar meson. In contrast, $\xi\neq0$ suppresses pseudoscalar-meson two-body decays, and pseudoscalar-meson decays into a sterile neutrino, a charged lepton, and a vector meson.
The interference terms in the meson decay rates between the Wilson coefficients $C_{\rm VRR}^{(6)}$ and $C_{\rm VLR}^{(6)}$ that appear for $\xi\neq0$ cause these effects.
For semi-leptonic $B$- and $D$-meson decays, the relevant hadronic matrix elements are
\begin{align}\label{eq:HMEBDtoP}
    \bra{h_{\rm PS}}\overline{q}_1\gamma^\mu P_{L,R}q_2\ket{B,D}&=+\frac{1}{2}\bra{h_{\rm PS}}\overline{q}_1\gamma^\mu q_2\ket{B,D},  \\
    \label{eq:HMEBDtoV}
    \bra{h_{\rm V}}\overline{q}_1\gamma^\mu  P_{L,R}q_2\ket{B,D}&=\mp\frac{1}{2}\bra{h_{\rm V}}\overline{q}_1\gamma^\mu \gamma^5 q_2\ket{B,D},
\end{align}
where $h_{\rm PS}$ and $h_V$ are pseudoscalar mesons and vector mesons, respectively, and the minus (plus) sign on the right-hand side of Eq.~\eqref{eq:HMEBDtoV} corresponds to the projection operator $P_L$ ($P_R$).
This relative minus sign in Eq.~\eqref{eq:HMEBDtoV} -- which also appears in the two-body leptonic decays -- leads to destructive interference terms linear in $\xi$.
These effects are sizable for large $\xi$.
Explicitly, the decay rates will be proportional to a factor $|C_{\rm VRR}^{(6)}\mp C_{\rm VLR1}^{(6)}|^2$.
In the limit where the CP-violating phase $\alpha=0$, we have $C_{\rm VLR1}^{(6)}=\frac{2\xi}{1+\xi^2} C_{\rm VRR}^{(6)}$, cf.~Eqs.~\eqref{eq:C6CCnuSMEFT}, \eqref{eq:CVLR12} and~\eqref{eq:CVRR}.
Hence, for values of $\xi$ approaching 1, the three-body decays into a pseudoscalar meson, and the two-body leptonic decays are suppressed.  

In type-I seesaw scenarios, sterile-neutrino production can occur via contributions from active-sterile neutrino mixing \textit{and} RH currents.
In the vanishing $W_L$-$W_R$ mixing limit, one can analytically determine under what conditions contributions from either the RH current or the active-sterile mixing dominate the meson-decay processes.
If only the lightest sterile neutrino $N_4$ is kinematically available, both contributions are equivalent when \begin{align}
\label{eq:RHvsMixingtype1}
    \left(\frac{M_{W_R}}{80 \text{ TeV}}\right)^4\left(\frac{m_1}{\text{meV}}\right)\left(\frac{\text{GeV}}{M_4}\right)\simeq 1\,.
\end{align}
For instance, given the lightest sterile neutrino at the pion threshold ($M_4=0.14$ GeV), the RH-current contributions will dominate for $M_{W_R}\leq 50$ TeV if $m_1=1$ meV.
Employing the relation above, we see that for relatively light $M_{W_R}< 10$ TeV and $\xi=0$, type-I seesaw scenarios converge to type-II scenarios since the minimal-mixing contributions are negligible for every reasonable choice of the lightest active-neutrino mass\footnote{This requires $\sum\limits_{i=1,2,3}m_i<0.12$ eV, considering cosmological constraints~\cite{Giusarma:2016phn, Vagnozzi:2017ovm,Vagnozzi:2018jhn,Tanseri:2022zfe,Planck:2018vyg}.} in the sterile-neutrino mass range $m_\pi<M_4<m_{B_s}$.

%% file: Sections/5_SterileDecay.tex
% !TEX root = ../Main.tex
\section{Sterile-neutrino decay}
\label{sec:Ndecay}

In this section, we study the leptonic and semi-leptonic decays of sterile neutrinos.
The leptonic decays may proceed via charged or neutral weak currents.
For the sterile-neutrino decays into mesons, we consider the pseudoscalar mesons $(\pi, K,\eta,\eta^\prime, D, D_s,\eta_c)$ and the vector mesons $(\rho,\omega, K^*,\phi, D^*, D_s^*)$.
We employ the physical parameters, decay constants, and form factors listed in Ref.~\cite{DeVries:2020jbs} to compute the relevant decay processes.

The decays into multi-meson final states warrant some discussion.
As argued in Ref.~\cite{Bondarenko:2018ptm}, these become relevant for sterile neutrinos heavier than roughly $1$ GeV in the minimal case. 
In the previous work, to avoid the tedious process of summing exclusive multi-meson decay channels, the contributions of multi-meson final states have been approximated by multiplying the decay width of sterile neutrinos into spectator quarks with an appropriate loop correction~\cite{Perl:1991gd, Braaten:1991qm, Gorishnii:1990vf}, which was initially employed to calculate multi-meson corrections in inclusive hadronic decay rates of the $\tau$ lepton.
We employ this procedure here as well for the minimal scenario.
For sterile neutrinos with masses below $1$ GeV, we compute the lifetimes by explicitly summing over final states that are purely leptonic or contain a single meson~\cite{Bondarenko:2018ptm, DeVries:2020jbs}, while for masses above $1$ GeV, we use 
\begin{align}\label{eq:hadrCor}
\Gamma(N_4\to \ell^- / \nu_\ell +\text{hadrons}) \simeq \left[1+\Delta_\text{QCD}(M_4)\right] {\Gamma_\text{tree}(N_4\to 
\ell^- /\nu_\ell + \overline{q_1}q_2)},
\end{align} where 
\begin{align}
    \Delta_{\rm QCD}(M_4)=\frac{\alpha_s(M_4)}{\pi}+5.5\frac{\alpha_s(M_4)^2}{\pi^2}+26.4\frac{\alpha_s(M_4)^3}{\pi^3}\,.
\end{align}

\subsection{Left-right symmetric models}

In mLRSM scenarios, determining the precision of the multi-meson corrections is more subtle.
In a type-II seesaw scenario without $W_L$-$W_R$ mixing, all mesons produced in sterile-neutrino decays couple to $W_R^\pm$ gauge bosons, whereas in the minimal scenario, they couple to $W_L^\pm$ gauge bosons.
Since QCD is invariant under parity transformations, we expect the multi-meson corrections to 
hold similarly in these scenarios.
In this case, employing Eq.~\eqref{eq:hadrCor} would be appropriate for sterile-neutrino masses above 1 GeV.
In principle, this is no longer true once we consider non-zero $W_L$-$W_R$ mixing scenarios since the $\xi$ dependence is dictated by the parity of the hadronic final state [in analogy to Eqs.~\eqref{eq:HMEBDtoP} and \eqref{eq:HMEBDtoV}].
Because $\xi<0.8$ cannot be too large in practice \cite{Maiezza:2010ic}, we neglect this difference and use Eq.~\eqref{eq:hadrCor} also when including $W_L$-$W_R$ mixing.
We do not expect this approximation to significantly affect our results, but we stress that explicitly computing the QCD corrections would be interesting for accounting for these $W_L$-$W_R$ mixing effects.

\begin{figure}
    \centering
    \includegraphics[width=0.7\textwidth]{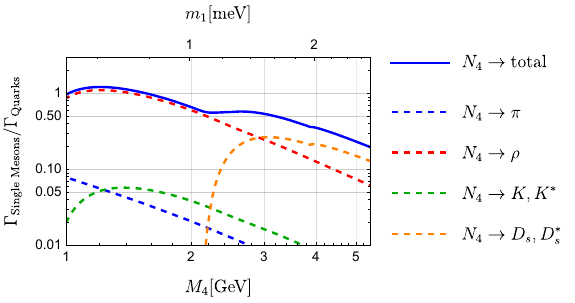}
    \caption{Ratios of the sterile-neutrino decay widths into individual mesons to the inclusive hadronic decay width calculated with Eq.~\eqref{eq:hadrCor} for $\xi=0.3$ and $M_6=100$ GeV, as functions of $M_4$/$m_1$.
    The solid line labeled with ``$N_4\to\text{total}$'' corresponds to the sum of the dashed curves.
    }
    \label{fig:singleMesonVSquarksxi03}
\end{figure}

In Fig.~\ref{fig:singleMesonVSquarksxi03}, we compare single-meson and multi-meson decay widths by naively applying the inclusive hadronic corrections in a scenario where $\xi=0.3$. 
The naive corrections appear significant for $M_4>1.55$ GeV, 
and assuming single-meson currents, as shown in the plot, 
leads to a systematic underestimate of the hadronic contributions to sterile-neutrino decays. 
Hence, for $\xi=0.3$, we calculate the total sterile-neutrino decay rates in this scenario by applying Eq.~\eqref{eq:hadrCor} for $1.55 \text{ GeV}<M_4<m_{B_s}$.
One can straightforwardly generalize this approach for different values of $\xi$.

\begin{figure}[t!]
    \centering
    \includegraphics[width=0.7\textwidth]{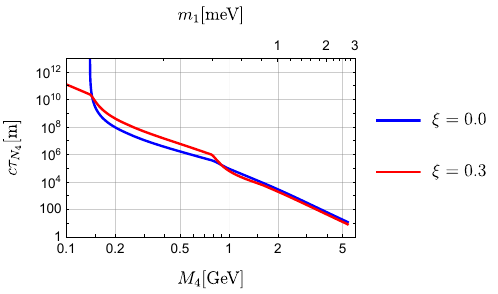}
    \caption{Proper decay lengths of sterile neutrinos in type-II seesaw mLRSM scenarios with and without $W_L$-$W_R$ mixing for $M_{W_R}=7$ TeV and $M_6=100$ GeV.
}
    \label{fig:LifeTimexi03}
\end{figure}

Fig.~\ref{fig:LifeTimexi03} depicts the proper decay length of the lightest sterile neutrino $c\tau_{N_4}$ for type-II 
seesaw scenarios with and without $W_L$-$W_R$ mixing, where $c$ is the speed of light and $\tau_{N_4}$ is the proper lifetime of the sterile neutrino.
Working in the NH -- which is what we will do throughout this work -- we consider a scenario with  $M_{W_R}=
7\,$TeV, $M_6=100\,$GeV such that $N_5$ is kinematically unavailable, and $\xi=0.3$.
It is perhaps surprising that around $M_4=1$\,GeV, the $N_4$ lifetimes are nearly equal despite the substantial 
interference effects caused by the non-zero $W_L$-$W_R$ mixing, but this is partially the case because 
we apply the same 
naive inclusive hadronic correction in both cases. 
These decay lengths allow for the 
detection of sterile neutrinos in prospective far-detector 
experiments at the \texttt{LHC}, as well as at \texttt{DUNE}, 
and \texttt{SHiP}. We discuss this further in 
Sec.~\ref{sec:simulations}.

\begin{figure}[t!]
    \centering
    \includegraphics[width=0.8\textwidth]{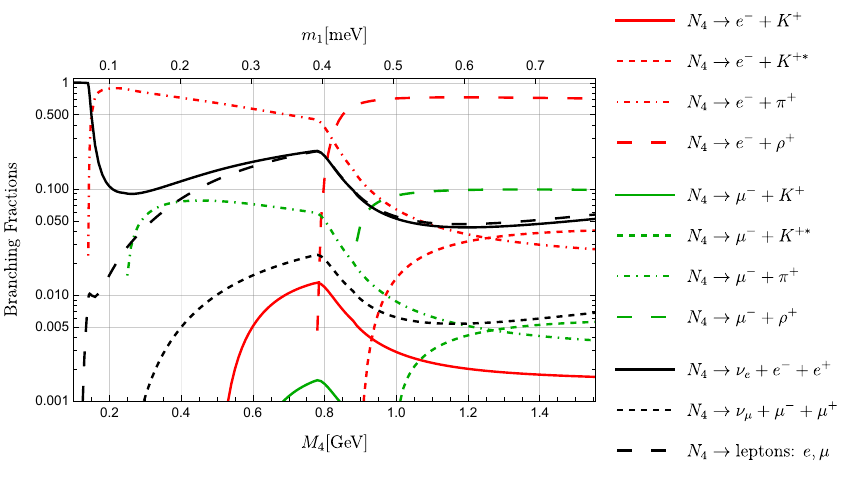}
    \caption{Sterile-neutrino decay branching fractions in a type-II seesaw mLRSM scenario where $\xi=0.3$ and $M_6=100$ GeV. The $N_4\to \text{leptons}+e,\mu$ branching fraction includes every $N_4$ decay mode that includes one electron and one muon.
    }
    \label{fig:BrFracxi03}
\end{figure}

In Fig.~\ref{fig:BrFracxi03}, the primary decay branching ratios of the lightest sterile neutrino $N_4$ into SM leptons and bound-state mesons are depicted for 
$M_4<1.55\,$GeV.
The hadronic decays dominate the sterile-neutrino decay processes once they become kinematically accessible around the charged pion threshold $M_4>M_{\pi^\pm}=139.6\,$MeV.
The vector meson $N_4\to e^-/\mu^- + \rho^+$ decay modes dominate at larger $M_4$, which is unsurprising as a result of the constructive interference terms in the decay rates being linear in $\xi$ and the fact that these processes are Cabibbo-allowed.
These branching fractions are independent of the charged RH gauge-boson mass $M_{W_R}$ since all decay rates scale as $M_{W_R}^{-4}$.

\begin{figure}[t]
    \centering
    \includegraphics[width=\textwidth]{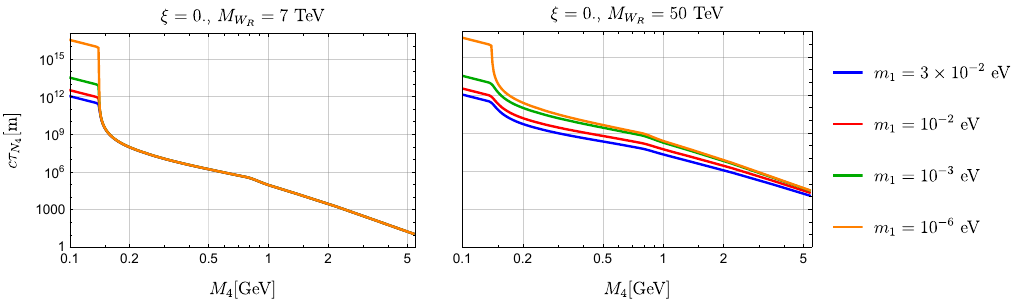}
    \vspace*{-0.6cm}
    \caption{Proper decay lengths of the lightest sterile neutrino in type-I seesaw scenarios for several values of $m_1$ and $M_{W_R}$, with $\xi=0$.}
    \label{fig:Type17vs30TeV}
\end{figure}

In Fig.~\ref{fig:Type17vs30TeV}, the proper decay lengths for the lightest sterile neutrino $N_4$ as functions of $M_4$ are shown in type-I seesaw scenarios without $W_L$-$W_R$ mixing for various values of the lightest active neutrino's mass $m_1$ and the RH gauge boson's mass $M_{W_R}$.
The maximum value $m_1=3\times10^{-2}\,$eV is close to the upper limit permissible under the cosmological constraints on the active-neutrino masses~\cite{Giusarma:2016phn, Vagnozzi:2017ovm,Vagnozzi:2018jhn,Tanseri:2022zfe,Planck:2018vyg}.
The left (right) panel shows the scenario where $M_{W_R}=7\,$TeV ($50\,$TeV).
For $M_{W_R}=7\,$TeV, when the sterile-neutrino mass surpasses the charged-pion threshold, and RH currents start contributing to the sterile-neutrino decays, the dependence of the sterile-neutrino lifetime on the lightest sterile-neutrino mass immediately diminishes.
This is due to the relatively light mass of the charged RH gauge boson.
For $M_4<M_{\pi^\pm}$, the only kinematically available decay channels are neutral-current decays, which depend linearly on $m_1$.
In contrast, the sterile-neutrino lifetimes for $M_{W_R}=50\,$TeV exhibit a more substantial dependence on $m_1$ across the relevant $M_4$ mass range.
This dependence becomes progressively less significant as $M_4$ increases.
The proper-lifetime behavior is similar in a scenario where $\xi=0.3$.

%% file: Sections/6_Neutrinoless.tex
% !TEX root = ../Main.tex
\section{Neutrinoless double beta decay}
\label{sec:0vbb}

Neutrinos being Majorana states and lepton number violation are key features of the mLRSM.
These features lead to non-zero $0\nu\beta\beta$ decay rates.
In addition to the usual contributions arising from the exchange of light active neutrinos between two nucleons in a nucleus, interesting additional contributions appear in the mLRSM. 

First of all, with just LH weak interactions, additional contributions arise from the exchange of the heavier sterile neutrinos as a result of the mixing between $\nu_L$ and $\nu_R$.
Then, in the mLRSM, owing to the RH gauge interactions, there are contributions solely from $\nu_R$ exchange.
The $W_L$-$W_R$ mixing also leads to additional topologies.
Further, there can be diagrams involving minimal mixing and right-handed gauge interactions (sometimes these are called the $\lambda$ and $\eta$ mechanisms~\cite{Doi:1985dx}).
Finally, there are contributions from the exchange of the doubly charged scalars $\delta_{L,R}^{++}$. However, in scenarios with relatively light RH neutrinos, these contributions are sub-leading and can be neglected \cite{deVries:2022nyh}.
Which of these contributions dominates depends on the interplay between the sterile-neutrino masses, the $M_{W_L}/M_{W_R}$ mass ratio, and the amount of $W_L$-$W_R$ mixing.

The various contributions to $0\nu\beta\beta$ decays in the mLRSM have been investigated in different approaches~\cite{Doi:1985dx, Rodejohann:2011mu, Nemevsek:2011aa, BhupalDev:2014qbx, Li:2020flq}.
In the last years, significant progress has been made towards a systematic EFT description of the $0\nu\beta\beta$ decays that takes into account the large-scale separations between the lepton-number-violating mechanism, the typical scale of nuclear physics, and the energy release in the $0\nu\beta\beta$ decays \cite{Prezeau:2003xn, Cirigliano:2017djv, Cirigliano:2018yza, Graf:2018ozy, Deppisch:2020ztt}.
This framework has been used to compute the $0\nu\beta\beta$ decay rates within the mLRSM.
The formulae are rather lengthy and explicitly given in Ref.~\cite{deVries:2022nyh}. Here, we 
discuss the salient features.

Within the mLRSM, the $0\nu\beta\beta$ decay amplitude mainly depends on two structures\footnote{In this section we sometimes use for notational convenience $m_{1,2,3,4,5,6}$ instead of $m_{1,2,3}$ and $M_{4,5,6}$.}
\begin{equation}
\mathcal A = \frac{g_A^2 G_F^2 m_e}{\pi R_A}\left[ \sum_{i=1}^6 \mathcal A_L(m_i)\,\bar e(k_1) P_R e^c(k_2) + \sum_{i=1}^6 \mathcal A_R(m_i)\,\bar e(k_1) P_L e^c(k_2)\right]+\dots\,,
\end{equation}
where $g_A \simeq 1.27$ is the nucleon axial charge and $R_A \simeq 1.2 \,A^{1/3}\,\mathrm{fm}$ is the nuclear radius of a nucleus with $A$ nucleons. The
dots denote other structures that are sub-leading~\cite{deVries:2022nyh}.
After integrating over the final state phase space, the $0\nu\beta\beta$ decay rate is then given by
\begin{equation}\label{eq:T1/2}
\left(T^{0\nu}_{1/2}\right)^{-1} = g_A^4 \left[ G_{01}  \left( |\mathcal A_{L}|^2 + |\mathcal A_{R}|^2 \right)
- 2 (G_{01} - G_{04}) \textrm{Re} \mathcal A_{L}^* \mathcal A_{R} \right]\,,
\end{equation}
where $\mathcal A_{L,R}=\sum_{i=1}^6 \mathcal A_{L,R}(m_i)$ and $G_{01,04}$ are atomic phase-space factors.
We will consider ${}^{136}$Xe for which $\{G_{01},\,G_{04}\} =\{1.5,\,3.2\}\cdot 10^{-14}\,\mathrm{yr}^{-1}$. 

The structure $A_L(m_i)$ involves the LH charged leptonic current and includes the exchange of the active Majorana neutrinos.
The relevant expression is given by
\begin{equation}\label{AL}
\mathcal A_L(m_i) = -\frac{m_i}{4 m_e} \mathcal M(m_i)\,\left(C_{\mathrm{VLL}}^{(6)}\right)^2_{udei}\,,
\end{equation}
in terms of the nuclear matrix element (NME) $\mathcal M(m_i)$, that depends on the mass of the exchanged neutrino (technically, it is a combination of hadronic and nuclear matrix elements, but we use NME here for simplicity).
In the modern approach, the NMEs are computed by combining chiral EFT with many-body nuclear methods.
For a detailed discussion of the mass dependence of the NMEs, we refer to Refs.~\cite{Dekens:2020ttz, Dekens:2023iyc, Dekens:2024hlz}.

We recall that in the mLRSM$\big(C_{\mathrm{VLL}}^{(6)}\big)_{udei} = - 2 V_{ud} (PU)_{ei}$.
In the type-II seesaw limit, the $6\times6$ $U$-matrix is block-diagonal, and the LH amplitude only sums over the active neutrinos.
In this case, we can neglect the neutrino-mass dependence of the NMEs and obtain
\begin{equation}
\mathcal A_L \simeq  -\frac{ V_{ud}^2}{ m_e} \mathcal{M}(0) \sum_{i=1}^3  
\left(U_{\mathrm{PMNS}}\right)_{ei}^2 m_i \equiv  -\frac{ V_{ud}^2}{ m_e} \mathcal{M}(0)\,m_{\beta\beta}\,,
\end{equation}
in terms of the effective Majorana neutrino mass $m_{\beta\beta}$ and $\mathcal{M}(0) = \mathcal O(1)$.
Beyond the type-II limit, $\mathcal A_L(m_i)$ generally depends on the exchange of sterile neutrinos.

The second structure $A_R(m_i)$ involves RH charged lepton currents.
Within the mLRSM, it consists of two contributions 
\begin{equation}
\mathcal A_R(m_i) = -\frac{m_i}{4 m_e} \left[\mathcal M(m_i)\,\left(C_{\mathrm{VRR}}^{(6)}+C_{\mathrm{VLR}}^{(6)}\right)^2_{udei} + 2 \mathcal M_{LR}(m_i)\,\left(C_{\mathrm{VRR}}^{(6)}\right)_{udei}\left(C_{\mathrm{VLR}}^{(6)}\right)_{udei}\right]\,.
\end{equation}
The first term corresponds to two RH charged quark currents.
Because QCD conserves parity, the associated NME is the same as
in Eq.~\eqref{AL}. The second term involves both a left- and a 
right-handed charged quark current and comes with its own NME.

In the type-II limit and turning off $W_L$-$W_R$ mixing (setting $\xi =0$ and thus $C_{\text{VLR}}^{(6)} =0 $), we obtain 
\begin{equation}
\mathcal A^{\xi=0}_R \simeq  -\frac{ V_{ud}^2}{ m_e} \frac{M^4_{W_L}}{M^4_{W_R}}\sum_{i=4}^6 \mathcal M(M_i) \left(U_{\mathrm{PMNS}}\right)_{e I}^2\,M_i\,,
\end{equation}
where $I \equiv i-3$.
We can now compare the RH contributions to the standard LH contributions in Eq.~\eqref{AL}.
For very light sterile neutrinos $M_{4,5,6} \ll m_\pi$ (not the focus of this work), we can again neglect the neutrino mass in the NME and obtain
\begin{equation}
\frac{\mathcal A_R^{\xi=0}}{\mathcal A_L} \sim \frac{M^4_{W_L}}{M^4_{W_R}} \frac{M_{\beta\beta}}{m_{\beta\beta}} \simeq (4 \cdot 10^3) \left(\frac{1\,\mathrm{TeV}}{M_{W_R}}\right)^4 \left(\frac{0.01\,\mathrm{eV}}{m_{\beta\beta}}\right) \left(\frac{M_{\beta\beta}}{1\,\mathrm{MeV}}\right)\,,
\end{equation}
where $M_{\beta\beta} \equiv \sum_{i=4}^6 \left(U_{\mathrm{PMNS}}\right)_{e I}^2\,M_i$.
For example, for $M_{\beta\beta} \simeq 1$ MeV and $m_{\beta\beta} \simeq 0.1$ eV, RH contributions dominate for $M_{W_R} \leq 8$ TeV. 

In the opposite limit $M_{4,5,6} \gg m_\pi$, we have $\mathcal{M}(M_i) \sim m_\pi^2/M_i^2$ so that $A_R$ is dominated by the lightest sterile neutrino $M_4$.
In this limit, we have roughly 
\begin{equation}
\frac{\mathcal A^{\xi=0}_R}{\mathcal A_L} \sim \frac{M^4_{W_L}}{M^4_{W_R}} \frac{m_\pi^2}{m_{\beta\beta} M_4} \simeq (8 \cdot 10^4) \left(\frac{1\,\mathrm{TeV}}{M_{W_R}}\right)^4 \left(\frac{0.01\,\mathrm{eV}}{m_{\beta\beta}}\right) \left(\frac{1\,\mathrm{GeV}}{M_4}\right)\,.
\end{equation}
For $M_4= 1\,\mathrm{GeV}\ll M_{5,6}$ and $m_{\beta\beta}=0.1$ eV, the RH contributions dominate for $M_{W_R} \leq 20$ TeV.
While these expressions are only estimates, they give an idea of the scale that can be reached.

The situation becomes more complicated and interesting once we turn on $W_L$-$W_R$ mixing.
This leads to a non-zero value of $C_{\mathrm{VLR}}$ and an additional contribution to $\mathcal A_R$. In the type-II limit, we can write 
\begin{equation}
\mathcal A^{\xi}_R \simeq  \mathcal A^{\xi=0}\left[1+ \frac{4\xi^2}{(1+\xi^2)^2}e^{-2i\alpha}\right] - \frac{V_{ud}^2}{m_e} \frac{M^4_{W_L}}{M^4_{W_R}} \frac{4 \xi e^{-i \alpha}}{1+\xi^2}\sum_{i=4}^6 \mathcal M_{LR}(M_i) \left(U_{\mathrm{PMNS}}\right)_{e I}^2\,M_i\,.
\end{equation}
Because $\xi <0.8$ \cite{Maiezza:2010ic} cannot be too large, the change in $\mathcal A^{\xi=0}$ in the first term is not too severe.
The second term, however, can dominate $\mathcal A^{\xi}_R$.
In particular, for $M_i \geq 1$ GeV, we can integrate out the sterile neutrinos, and the $W_L$-$W_R$ mixing leads to dimension-nine lepton-number-violating operators of the form $(\bar u_L \gamma^\mu d_L)\,(\bar u_R \gamma^\mu d_R)\,\bar e_L C \bar e_L^T$.
This operator has the right chiral properties to induce, at lower energies, processes of the form $\pi^-\pi^- \rightarrow e^- e^-$, which lead to enhanced $0\nu\beta\beta$ amplitudes once the pions are exchanged between nucleons in a nucleus~\cite{Prezeau:2003xn, 
 Cirigliano:2018yza, Nicholson:2018mwc}.
This enhancement disappears for smaller $M_i \leq m_\pi$ but is still present for the 
sterile-neutrino masses we are interested in. 

\begin{figure}[t!]
    \centering
    \includegraphics[width=\textwidth]{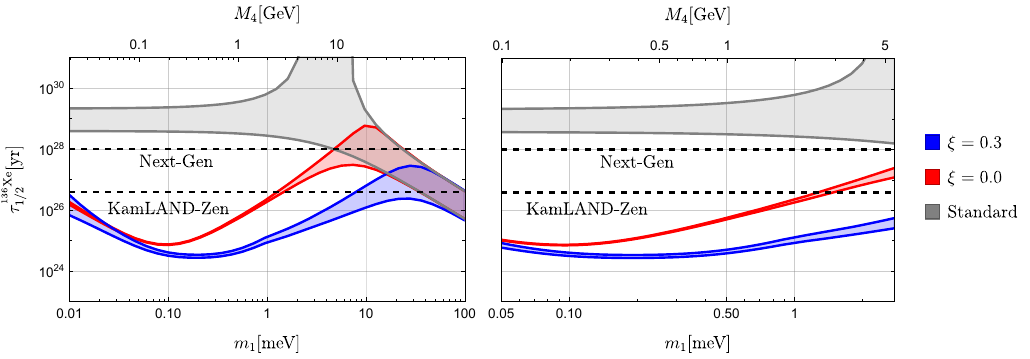}
    \vspace*{-0.6cm}
    \caption{$0\nu\beta\beta$ lifetime of Xe${}^{136}$ in the type-II mLRSM in the NH as a function of the lightest (sterile) neutrino mass. We have set the heaviest sterile-neutrino mass at 100 GeV and $M_{W_R}=10$ TeV and $\xi=0.3$ (blue) and $\xi=0$ (red). The gray regions correspond to the standard mechanism associated with the exchange of light active Majorana neutrinos without right-handed interactions. The right panel is the same as the left but zoomed in on the mass range most relevant for this work.}
    \label{fig:0vbb}
\end{figure}

To illustrate the constraints that can be set from $0\nu\beta\beta$ experiments, we show the $0\nu\beta\beta$ lifetime of ${}^{136}$Xe in Fig.~\ref{fig:0vbb} in the type-II limit in the NH as a function of the lightest neutrino mass $m_1$ (or equivalently, the lightest sterile-neutrino mass $M_4$). 
We have set $M_6 = 100$\,GeV and $M_{W_R}=10$ TeV for concreteness and considered two cases for the $W_L$-$W_R$ mixing parameters: $\xi=0$ (red) and $\xi=0.3$ (blue). 
The gray regions in both figures indicate the lifetime in the case of just three light Majorana neutrinos without contributions from sterile neutrinos or RH currents (we set $M_{W_R} \rightarrow \infty$). 
For $m_1 <1$ meV, this leads to lifetimes of the order of $10^{29}$\,y (the band is caused by marginalizing the unknown Majorana phases), roughly three orders of magnitude longer than the current best limit from KamLAND-Zen, $T^{0\nu}_{1/2} > 3.8 \cdot 10^{26}$\,y \cite{KamLAND-Zen:2024eml}, shown by the dashed horizontal line.
We note that even the next-generation experiments, nEXO and LEGEND, projected to improve the sensitivities by a factor of $\sim 100$~\cite{nEXO:2017nam,LEGEND:2017cdu}, cannot reach that, as shown also with a dashed horizontal line.
In the mLRSM, the presence of relatively light sterile neutrinos that are charged under $SU(2)_R$, as indicated by the blue and red bands, can significantly reduce the lifetimes, reaching lifetimes around $10^{25}$\,y for HNLs with masses of $\mathcal O(10^2)$\,MeV. 
We obtain the width of the bands by marginalizing over the unknown Majorana phases and the phases $\alpha$ and $\theta_L$.
The effect of $W_L$-$W_R$ mixing can be severe in this mass range because of the chiral enhancement discussed above. 
In the right panel of Fig.~\ref{fig:0vbb}, we show the same results but now zoom in on the mass region for $M_4$ most relevant for displaced-vertex searches. 
We see that $0\nu\beta\beta$ limits rule out a large part of this parameter space even for a $M_{W_R}=10$\,TeV mass. 

\begin{figure}[t!]
    \centering
    \includegraphics[width=\textwidth]{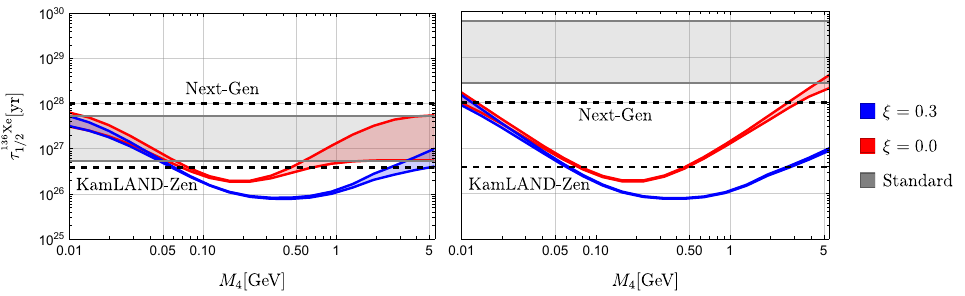}
    \vspace*{-0.6cm}
    \caption{$0\nu\beta\beta$ lifetime of Xe${}^{136}$ in the type-I mLRSM in the NH as a function of the lightest HNL. We have set $m_1=0.03$ eV (left) and $m_1=0.001$ eV (right), $M_5=M_6=100$ GeV, $M_{W_R}=15$ TeV,and $\xi=0.3$ (blue) and $\xi=0$ (red). The gray regions correspond to the standard mechanism associated with the exchange of light active Majorana neutrinos without RH interactions.}
    \label{fig:0vbbtype1}
\end{figure}

We can perform very similar calculations in the type-I scenario. In this case, we need to specify the lightest neutrino mass, the hierarchy, and the values for all HNLs to compute the $0\nu\beta\beta$ lifetime. In Fig.~\ref{fig:0vbbtype1}, we have considered a scenario with $M_{W_R} = 15$ TeV for the lightest HNL in the mass region of interest, and we assumed the other HNLs to be significantly heavier. The difference between the left and right panel is the value of the lightest neutrino mass ($m_1= 0.03$ eV left and $m_1=0.001$ eV right). As can be seen by comparing the two panels, the value of $m_1$ affects the $0\nu\beta\beta$ curves. Still, it barely influences the constraints on $m_{W_R}$ because RH currents dominate the decay rates at the points where the current experimental limit is saturated. 

%% file: Sections/7_ColliderAnalysis.tex
% !TEX root = ../Main.tex
\section{Collider and fixed-target analysis}
\label{sec:simulations}
This section briefly introduces our computation method of estimating the expected number of observed sterile-neutrino-decay events in the considered experiments.
We include the following \texttt{LHC} far-detector setups with the associated integrated luminosities: \texttt{ANUBIS} (\SI{3}{\atto\per\barn})~\cite{Bauer:2019vqk}, \texttt{FASER} (\SI{150}{\femto\per\barn}) and \texttt{FASER2} (\SI{3}{\atto\per\barn})~\cite{Feng:2017uoz,FASER:2018eoc}, \texttt{MATHUSLA} (\SI{3}{\atto\per\barn})~\cite{Chou:2016lxi,Curtin:2018mvb,MATHUSLA:2020uve}, \texttt{FACET} (\SI{3}{\atto\per\barn})~\cite{Cerci:2021nlb}, \texttt{CODEX-b} (\SI{300}{\femto\per\barn})~\cite{Gligorov:2017nwh}, and \texttt{MoEDAL-MAPP1} (\SI{30}{\femto\per\barn}) and \texttt{MoEDAL-MAPP2} (\SI{300}{\femto\per\barn})~\cite{Pinfold:2019nqj,Pinfold:2019zwp}.
For a more detailed summary of the \texttt{LHC} far-detector experiments, we refer to e.g.~Refs.~\cite{DeVries:2020jbs, Gunther:2023vmz} and the references therein.
We further investigate the \texttt{DUNE} ($1.1\times 10^{22}$ protons on target (POTs)) near detector of the \texttt{LBNF} at Fermilab with a total run time of $10$ years~\cite{DUNE:2020lwj, DUNE:2020ypp, DUNE:2020jqi, DUNE:2021cuw, DUNE:2021mtg}, and the recently approved \texttt{LHC} beam-dump experiment \texttt{SHiP} ($2\times 10^{20}$ POTs) with five years' operation duration~\cite{SHiP:2015vad, Alekhin:2015byh, SHiP:2018xqw, SHiP:2021nfo}.
We consider the case of a $120$-GeV ($400$-GeV) proton beam hitting a target at \texttt{DUNE} (\texttt{SHiP}).
The \texttt{DUNE} near detector~\cite{DUNE:2016evb, DUNE:2016rla} employs a $194$ m long decay pipe with a radius of $2$ m placed $27$ m downstream from the target, where the kaons are supposed to decay.
Its near detector is constructed at a distance of $574$ m downstream from the target and has a length of $6.4$ m and a width of $3.5$ m.
The \texttt{SHiP} experiment~\cite{SHiP:2021nfo} should instrument the detector with a distance of $45$ m from the IP and a length of $50$ m.
With a pyramidal frustum shape, the detector has dimensions $1.5$ m $\times$ $4.3$ m ($5$ m $\times$ $10$ m) with its front (rear) surface.

Next, we will discuss the computation of the number of produced sterile neutrinos.
Considering sterile neutrinos originating from meson decays, $N_{\mathrm{prod}}$ is given by
\begin{equation}\label{eq:neutrino_production}
    N_{M,N}^{\mathrm{prod}} = N_M \cdot \mathrm{Br}\big(M\rightarrow N + X\big)\,,
\end{equation}
where $N_M$ is the number of mesons $M$ produced at an experiment for its total runtime, which we list in Tab.~\ref{tab:meson_numbers} for the \texttt{LHC}, \texttt{DUNE}, and \texttt{SHiP} experiments, and $\mathrm{Br}\big(M\rightarrow N + X\big)$ is the branching ratio of a meson $M$ into the final states that include a sterile neutrino (see Sec.~\ref{sec:Nproduction}).

{\renewcommand*{\arraystretch}{1.1}
\begin{table}[t]
\footnotesize
    \centering
    \begin{tabular}[T]{|c|ccc|}
        \hline
        Meson $M$ & $N_M$ (\texttt{LHC}) & $N_M$ (\texttt{DUNE})& $N_M$ (\texttt{SHiP}) \\
        \hline
        $K^\pm$ & $2.38\times 10^{18}$ &  $5.76\times 10^{21}$   &$1.9\times 10^{19}$\\
        $K_{L}$ & $1.31\times 10^{18}$ & $5.76\times 10^{21}$    &$1.9\times 10^{19}$\\
        $K_{S}$ & $1.31\times 10^{18}$ & $5.76\times 10^{21}$    &$1.9\times 10^{19}$\\\hline
        $D^\pm$ & $2.04\times 10^{16}$ & $2.28\times 10^{17}$    &$1.4\times 10^{17}$\\
        $D^0$ & $3.89\times 10^{16}$ & $6.95\times 10^{17}$      &$4.3\times 10^{17}$\\
        $D_s^\pm $ & $6.62\times 10^{15}$ & $9.68\times 10^{16}$ &$6.0\times 10^{16}$\\ \hline
        $B^\pm$ & $1.46\times 10^{15}$ & $4.46\times 10^{11}$    &$2.7\times 10^{13}$ \\
        $B^0$ & $1.46\times 10^{15}$ & $4.46\times 10^{11}$      &$2.7\times 10^{13}$\\
        $B_s^0$ & $2.53\times 10^{14}$ & $1.11\times 10^{11}$    &$7.2\times 10^{12}$ \\
        \hline
    \end{tabular}
    \caption{Number of mesons $M$ over the full $4\pi$ solid angle~\cite{DeVries:2020jbs,Gunther:2023vmz,Krasnov:2019kdc,Bondarenko:2018ptm}.
    For the \texttt{LHC}, we show $N_M$ for a center-of-mass energy of \SI{14}{\tera\eV} and a total integrated luminosity of \SI{3}{\atto\per\barn}.
    For \texttt{DUNE} with a proton beam energy of \SI{120}{\giga\eV}, $N_M$ numbers are presented for a runtime of ten years with 1.1$\times 10^{22}$ POTs.
    For \texttt{SHiP} with a proton beam energy of \SI{400}{\giga\eV}, we list $N_M$ for a total operation time of 5 years with $2\times 10^{20}$ POTs.
    The superscript $\pm$ means the sum of the numbers of the positively charged and the negatively charged mesons.
    Similarly, $D^0$, $B^0$, and $B^0_s$ all denote the sum of the numbers of the two charge-conjugated states.
    }\label{tab:meson_numbers}
\end{table}
}

To evaluate the fraction of $N_{\mathrm{prod}}$ observed through sterile-neutrino decays in the detectors, we apply Monte-Carlo techniques to determine the geometrical acceptance for each detector setup.
For a single $N$ with speed $\beta_i$, boost factor $\gamma_i$ and proper lifetime $\tau_N$ (see Sec.~\ref{sec:Ndecay}) produced in the decay of a meson $M$, the probability of decaying inside the fiducial volume of a detector is given by
\begin{equation}
    P_{M,i}\big[N\text{ }\mathrm{in}\text{ }\mathrm{f.v.}\big]=\exp\bigg[-\frac{L_{T,i}}{\beta_i\gamma_i c\tau_N}\bigg]\bigg(1-\exp\bigg[-\frac{L_{I,i}}{\beta_i\gamma_i c\tau_N}\bigg]\bigg)\,.\label{eqn:decay_probability}
\end{equation}
Here, $L_{T,i}$ is the length $N_i$ travels to reach the detector, $L_{I,i}$ is the length $N_i$ would travel through the detector if it would not decay inside, and $c$ is the speed of light.
The values of $L_{T,i}$ and $L_{I,i}$ depend not only on the experiment on hand but also on the traveling direction of $N_i$ as well as the vertex at which the meson decays into $N_i$. 
To understand their typical values, we refer to Refs.~\cite{DeVries:2020jbs,Gunther:2023vmz} for a summary of the geometrical setup of the experiments considered in this work.
If $ N_i$ does not travel towards the detector, the decay probability inside the detector is zero.
The charm and bottom mesons can be legitimately assumed to be decaying at the IP.
However, we cannot do so for the kaons with their longer lifetimes.
We thus also consider their displaced decay positions within the simulation.
The computation should include only the kaons decaying before reaching areas of rock, shielding infrastructure, hadron absorbers, calorimeters, etc.
For \texttt{DUNE}, there is no such problem as a specific decay pipe is instrumented at the near-detector experiment, while for the far detectors at the \texttt{LHC}, care should be taken and we refer to Ref.~\cite{Gunther:2023vmz} for the detail.
For the \texttt{SHiP} experiment, the target is 1.4 m long~\cite{SHiP:2021nfo}, followed immediately by a hadron absorber and then muon shielding; thus, we require that the kaons decay before traveling one nuclear interaction length ($\sim 20$ cm) into the target.

Using the event generator \texttt{PYTHIA8}~\cite{Bierlich:2022pfr, Sjostrand:2014zea} in combination with a three-dimensional approximation of the fiducial volume~\cite{Gunther:2023vmz}\footnote{In the three-dimensional approximation we can include a meson traveling a macroscopic distance before decaying into the sterile neutrino. This allows us to include decays of the long-lived kaons in the analysis.}, we simulate $N_{\mathrm{MC}}=10^6$ proton-on-proton events with the corresponding beam-energy setups and calculate an average decay probability with
\begin{equation}
    \big<P_M\big[N_i \text{ }\mathrm{in}\text{ }\mathrm{f.v.}\big]\big> = \frac{1}{N_{\mathrm{MC}}}\sum_i^{N_{\mathrm{MC}}} P_{M,i}\big[N\text{ }\mathrm{in}\text{ }\mathrm{f.v.}\big].
\end{equation}

Lastly, the detector must be able to reconstruct the decay vertex of the sterile neutrino.
We call final states 'visible' if we expect that the detector setup is capable of vertex reconstruction for that given final state. 'Invisible' final states, such as a final state consisting of only active neutrinos, are excluded from the analysis. The expected number of observed sterile-neutrino decays is given by
\begin{equation}
    N_N^{\mathrm{obs}}=\mathrm{Br}\big(N\rightarrow \mathrm{visible}\big)\sum_M N_{M,N}^{\mathrm{prod}} \cdot \big<P_M\big[N\text{ }\mathrm{in}\text{ }\mathrm{f.v.}\big]\big>\cdot h_M^{\text{decay}}\,,
\end{equation} 
where $\mathrm{Br}\big(N\rightarrow \mathrm{visible}\big)$ is the branching ratio of the sterile neutrino into visible final states, and we assume 100\% detection efficiency.
$h_M^{\text{decay}}$ is the fraction of the mesons $M$ decaying in time, cf.~the discussion below Eq.~\eqref{eqn:decay_probability}.
It is, therefore, equal to one for the charm and bottom mesons but is non-zero for kaons at the \texttt{LHC} far detectors and the \texttt{SHiP} experiment.

%% file: Sections/8_NumericalResults.tex
% !TEX root = ../Main.tex
\section{Numerical results}
\label{sec:numericalResults}

In the following, we discuss the results of the numerical simulations as introduced in the previous sections. 

\subsection{Type-II seesaw benchmark scenarios}

\begin{figure}[t!]
    \centering
    \includegraphics[width=0.495\textwidth]{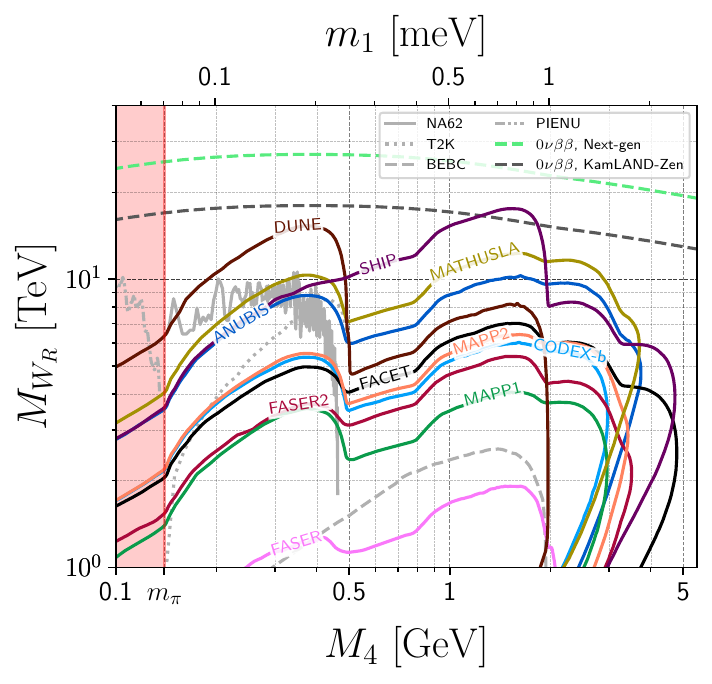}
    \includegraphics[width=0.495\textwidth]{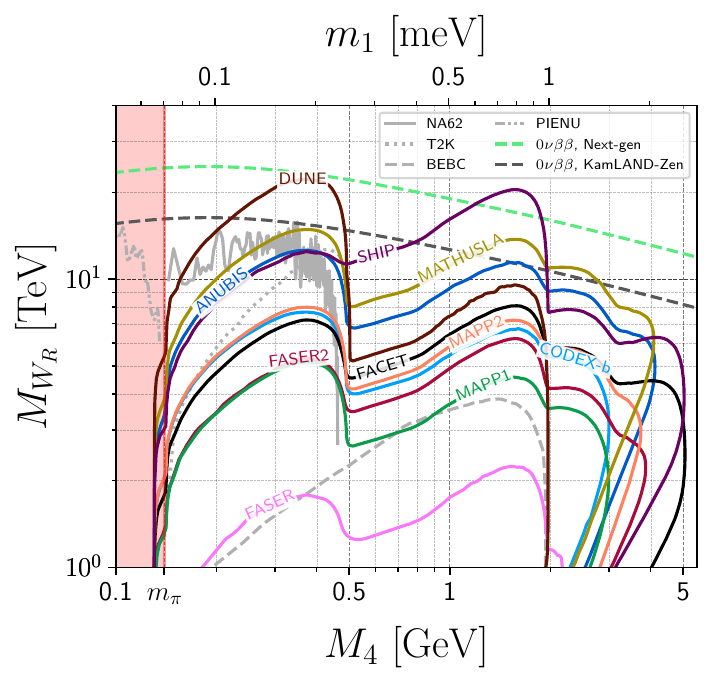}
    \vspace*{-0.6cm}
    \caption{Predicted sensitivity reach of the next-gen \ovbb experiments and the DV searches (at the \texttt{LHC} far detectors, \texttt{DUNE} near detector, and beam-dump experiment \texttt{SHiP}) in the $M_4-M_{W_R}$ plane in type-II seesaw scenarios. For the $W_L$-$W_R$ mixing, we have set $\xi=0.3$ ($\xi=0$) in the left (right) panel.
    Also displayed are existing bounds obtained by recasting the results from the \ovbb experiment \texttt{KamLAND-Zen}~\cite{KamLAND-Zen:2024eml} and the search for long-lived sterile neutrinos at the \texttt{PIENU}~\cite{PIENU:2019usb}, \texttt{NA62}~\cite{NA62:2020mcv}, \texttt{T2K}~\cite{T2K:2019jwa}, and \texttt{BEBC}~\cite{WA66:1985mfx,Barouki:2022bkt} experiments.
    The region for $M_4<m_\pi$ is shaded in light red, as we do not include sterile neutrinos from pion decays in our simulation.
    We have set the heaviest sterile neutrino mass $M_6=100$\,GeV.}
    \label{fig:LR22}
\end{figure}

We first analyze type-II seesaw scenarios with ($\xi=0.3$) and without ($\xi=0.0$) $W_L$-$W_R$ mixing.
In Fig.~\ref{fig:LR22}, we have drawn isocurves of $N_{N}^{\text{obs}}=3$ in the $M_4\!-\!M_{W_R}$ plane for the \texttt{LHC} far detectors, \texttt{DUNE} near detector, and \texttt{SHiP}, as introduced in Sec.~\ref{sec:simulations}. 
To simplify the presentation while obtaining the sensitivity reach to the RH gauge-boson mass, we assume a scenario where $M_{5,6}>m_{B_s}$, ensuring that only the lightest sterile neutrino $N_4$ is kinematically relevant.
We achieve this by selecting a sizable heaviest sterile-neutrino mass $M_6=100\,$GeV, cf.~Eq.~\eqref{eq:m1m4massRelations}.
Choosing a substantially larger $M_6=1$\,TeV would not significantly impact the results. 
Below the kaon threshold $M_4<M_{K^\pm}$, \texttt{DUNE} has the highest sensitivity, whereas for $M_{K}^\pm<M_4<M_{D}^\pm$ and $M_{D}^\pm<M_4<M_{B_s}$, \texttt{SHiP} and \texttt{MATHUSLA} are predicted to have the strongest constraining power among the considered experiments, respectively.
Furthermore, we display together the latest exclusion bounds of the \ovbb experiment \texttt{KamLAND-Zen}~\cite{KamLAND-Zen:2024eml} on the RH gauge-boson mass.
These limits were obtained by calculating the ${}^{136}\text{Xe}$ lifetime, as discussed in Sec.~\ref{sec:0vbb}.
Here, to set the most conservative limits on $M_{W_R}$, we have maximized the $^{136}\text{Xe}$ lifetime over the unknown phases $\lambda_{1,2}$ and $\theta_L$.
Additionally, we depict the sensitivity reach of next-gen ton-scale experiments that are projected to probe the $^{136}\text{Xe}$ lifetime up to $\mathcal{O}(10^{28})$ years~\cite{nEXO:2017nam,LEGEND:2017cdu}.
We note that for $M_4$ below the pion threshold, the strongest sensitivity reach should, in principle, be derived by considering pion decays.
We do not consider these decays in our simulation as our tool \texttt{PYTHIA8} is not well validated for forward pion production. Therefore, we shade the corresponding parameter region with a light red color.
Nevertheless, in this mass regime, the leading bounds on $|U_{e4}|^2$ were reported by the \texttt{PIENU} collaboration~\cite{PIENU:2019usb}.
We have reinterpreted these bounds in terms of our model and will display them in the numerical results.

The left panel of Fig.~\ref{fig:LR22} shows that the highest sensitivity reach to the RH gauge-boson masses for $0.1\,$GeV$<M_4<5\,$GeV will be provided by next-generation \ovbb searches for $\xi=0.3$.
In this scenario, one can probe RH gauge-boson effects up to $M_{W_R}\sim 25$\,TeV with these  $0\nu\beta\beta$ experiments.
Compared to the current \ovbb bounds on $M_{W_R}$, \texttt{DUNE} and \texttt{SHiP} are the only experiments that provide a competitive sensitivity reach, probing at most $M_{W_R}\!\sim\!15$\,TeV.
The right panel shows the same scenario but for $\xi=0.0$.
There, the DV searches can be competitive with next-generation \ovbb experiments.
Below the kaon threshold, around $0.3\,\text{ GeV}< M_4< M_K^\pm$, \texttt{DUNE} and future \ovbb experiments exhibit similar sensitivity reach to RH currents.
Above the kaon threshold, for $1.2 \text{ GeV}< M_4 < 1.9\,\text{ GeV}$, \texttt{SHiP} outperforms the projected next-generation \ovbb experiments. 

Below the kaon threshold, the most stringent upper limits on the minimal mixing angle $|U_{e4}|^2$ between the active and sterile neutrinos have been determined by the experiments \texttt{NA62}~\cite{NA62:2020mcv} and \texttt{T2K}~\cite{T2K:2019jwa}.\footnote{As mentioned above, in our numerical simulation we do not take into account the sterile neutrinos from pion decays which should dominate the sensitivity reach for $M_4<m_{\pi}$, where the strongest present bounds on $|U_{e4}|^2$ were obtained at the PIENU experiment~\cite{PIENU:2019usb}. These bounds have been reinterpreted here in terms of our scenario.}
We translate these upper bounds into lower bounds on the RH gauge-boson mass, which agree with previous assessments~\cite{Alves:2023znq}.
We can compare these limits to the \ovbb constraints. 
For $\xi=0.0$, current \texttt{NA62} and \texttt{T2K} limits are competitive with current \ovbb constraints, whereas, for larger $\xi$, the \ovbb limits are more stringent. 
Upper bounds on the minimal mixing angle have also been determined by experiments such as \texttt{CHARM}~\cite{CHARM:1985nku, Orloff:2002de}, \texttt{DELPHI}~\cite{DELPHI:1996qcc}, \texttt{PS191}~\cite{Bernardi:1987ek, Ruchayskiy:2011aa}, \texttt{BEBC}~\cite{WA66:1985mfx, Barouki:2022bkt}, and \texttt{NuTeV}~\cite{NuTeV:1999kej}.
Below the kaon threshold, these experiments' constraints are weaker than those of \texttt{NA62}~\cite{NA62:2020mcv} and \texttt{T2K}~\cite{T2K:2019jwa}, and we therefore do not show them.
However, \texttt{BEBC}~\cite{WA66:1985mfx, Barouki:2022bkt} does extend the limits to larger sterile-neutrino masses leading the bounds on $|U_{e4}|^2$ for $M_4$ between the kaon and charm-meson thresholds, and its bounds are shown as dashed light-gray curves.
They are much less stringent than the $0\nu\beta\beta$ limits.

We predict that the sensitivity reach of (future) DV searches to $M_{W_R}$ is stronger for $\xi=0.0$ than for $\xi=0.3$.
Further, it is no coincidence that the \texttt{DUNE} sensitivity reach is most significantly impacted by non-zero values of $\xi$, as at \texttt{DUNE} predominantly (pseudoscalar) pions and kaons are produced, which are suppressed by the non-zero $W_L$-$W_R$ mixing.
Finally, as could be deduced from Fig.~\ref{fig:0vbb} (right), the \ovbb sensitivity reach is increased in a scenario with non-zero $W_L$-$W_R$ mixing.

\subsection{Type-I seesaw benchmark scenarios}

Determining the sensitivity reaches of the considered experiments in type-I benchmark scenarios is only sensible when minimal-mixing contributions are significant compared to the RH-current contributions. 
Otherwise, these type-I seesaw scenarios converge to type-II seesaw scenarios, and we gain no new information.
We again assume a scenario where only the lightest sterile neutrino $N_4$ is kinematically available. 
We determine whether the minimal-mixing or RH-current contributions dominate the sterile-neutrino production rate with Eq.~\eqref{eq:RHvsMixingtype1}.  
For sterile-neutrino decays, we perform the same task numerically.

\begin{figure}[t!]
    \centering
    \includegraphics[width=0.8\textwidth]{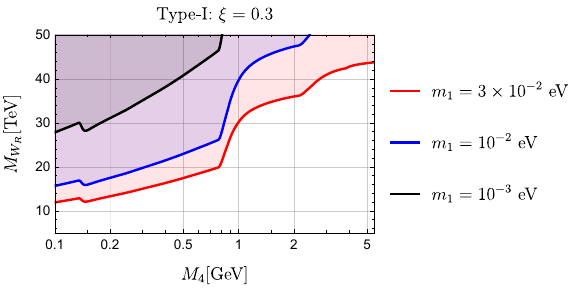}
    \caption{Isocurves in the $M_4$-$M_{W_R}$ plane, corresponding to equal contributions from RH-current and minimal-mixing effects to sterile-neutrino decays for several $m_1$ values in type-I seesaw scenarios.
    Below (above) each of the isocurves, the RH-current (minimal-mixing) effects dominate.}
    \label{fig:minmixvsRHdom}
\end{figure}

In Fig.~\ref{fig:minmixvsRHdom}, we show in the $M_4$-$M_{W_R}$ plane whether the minimal-mixing contributions or RH-current contributions dominate the sterile-neutrino decay for various $m_1$ values.
For example, we see that for a large $m_1=0.03$\,eV, the minimal-mixing contributions dominate the decays of the lightest sterile neutrinos for $M_4<0.9$\,GeV (around the $\rho$-meson threshold) if $M_{W_R}\gtrsim 20$\,TeV. 
As shown in Fig.~\ref{fig:LR22}, DV-search experiments such as \texttt{DUNE}, \texttt{SHiP}, and \texttt{MATHUSLA} 
can probe RH gauge-boson effects in this mass range in type-II seesaw scenarios.
This means minimal-mixing contributions can be significant for a relatively large $m_1$. 
Hence, we deem it interesting to analyze further the sensitivity reach in type-I seesaw benchmark scenarios, which we 
now do.

\begin{figure}[t!]
    \centering
    \includegraphics[width=0.495\textwidth]{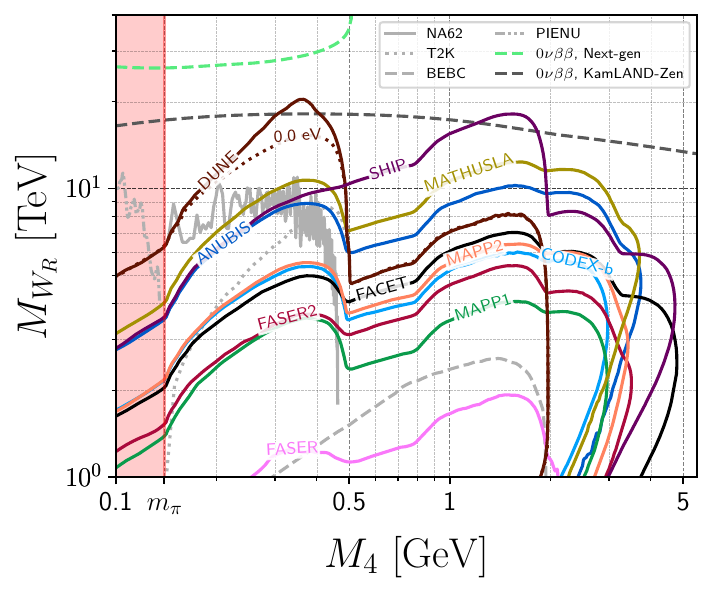}
    \includegraphics[width=0.495\textwidth]{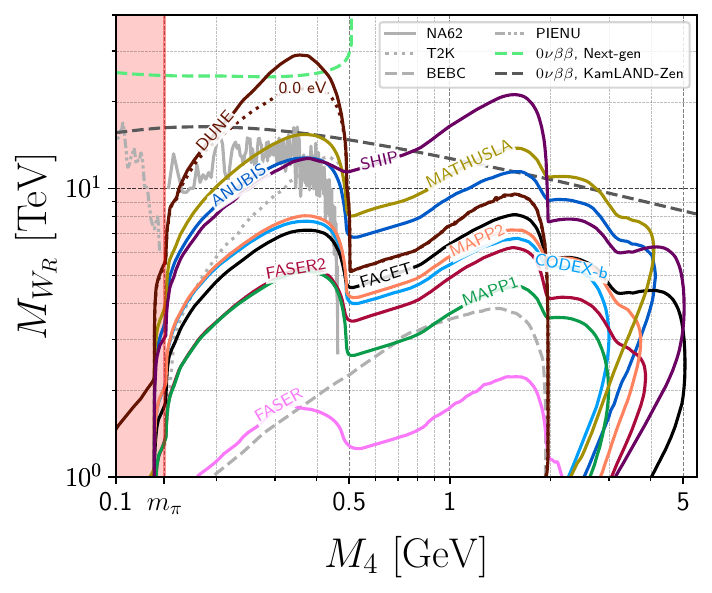}
    \vspace*{-0.6cm}
    \caption{Predicted sensitivity reach of the next-gen \ovbb experiments and the DV searches in the $M_4-M_{W_R}$ plane in the type-I seesaw scenario setting $m_1 =0.03$ eV. For the $W_L$-$W_R$ mixing, we have set $\xi=0.3$ ($\xi=0.0$) in the left (right) panel.
    We also show existing bounds from the \ovbb experiment KamLAND-Zen~\cite{KamLAND-Zen:2022tow}.
    We show recast bounds obtained from \texttt{PIENU}~\cite{PIENU:2019usb}, \texttt{NA62}~\cite{NA62:2020mcv}, \texttt{T2K}~\cite{T2K:2019jwa}, and \texttt{BEBC}~\cite{WA66:1985mfx,Barouki:2022bkt}.
    }
    \label{fig:LR12}
\end{figure}

To maximize the impact of minimal-mixing effects, we set $m_1=0.03$\,eV. In Fig.~\ref{fig:LR12}, we show isocurves 
where $N_N^{\text{obs}}=3$ for the (future) DV-search experiments in the considered type-I seesaw scenario. All the 
\texttt{LHC} far detectors that probe $\sim 10$\,TeV RH gauge-boson masses exhibit similar 
isocurves compared to 
the type-II scenario. This is unsurprising as RH contributions dominate at this magnitude of $M_{W_R}$. However, in the case 
of \texttt{DUNE}, the sensitivity in the type-I seesaw scenario is perceptibly higher than in the type-II seesaw scenario. Finally, we note that the results are insensitive to the choice between generalized $\mathcal{P}$-symmetry and $\mathcal{C}$-symmetry scenarios.

Since there are no direct relations between the active- and sterile-neutrino masses in the type-I seesaw scenario, there is less predictive power for the \ovbb sensitivity reach compared to type-II seesaw scenarios.
Nevertheless, comparing the \ovbb sensitivity reach to the far-detector sensitivity reaches is still interesting.
We choose the sterile-neutrino masses to be large ($M_{5,6}=100 \text{ GeV}\gg M_{B_s}$) to be conservative.
In Fig.~\ref{fig:LR12}, we show the results for current and next-gen experiments in this type-I seesaw scenario.
Again, the existing $0\nu\beta\beta$ constraints are stringent, but \texttt{DUNE} and \texttt{SHiP} will be more sensitive in parts of the parameter space.

Because of the relatively large $m_1$ at 0.03 eV, the minimal-mixing contributions are sufficiently large to be detected in next-generation $0\nu\beta\beta$ experiments for $M_4>0.5$\,GeV.
This is true irrespective of the mass of the RH gauge boson, and therefore, the green dashed line becomes vertical.
However, For $M_4<0.5$\,GeV, the active-neutrino contributions to the ${}^{136}\text{Xe}$ decay rate can cancel the sterile-neutrino contributions, analogous to the phase cancellations occurring in parts of the gray regions in Fig~\ref{fig:0vbb}.
In this case, the next-generation $0\nu\beta\beta$ experiments will be sensitive to $M_{W_R} = 25$\,TeV roughly.

%% file: Sections/9_Conclusions.tex
% !TEX root = ../Main.tex
\section{Conclusions}
\label{sec:conclusions}

Sterile neutrinos are prime candidates for explaining the lightness of the non-zero active neutrino masses
via the seesaw mechanism. 
These elusive particles could also address other major issues in particle physics and cosmology, 
such as the baryon asymmetry of the Universe and dark matter. Relatively light, GeV-scale sterile 
neutrinos are particularly interesting because they can be copiously produced at numerous current 
and future particle-physics experiments. This 
work investigates the phenomenology of GeV-scale sterile neutrinos within the mLRSM. There, sterile neutrinos couple to SM particles through minimal mixing with active neutrinos and via 
heavy right-handed gauge bosons induced by a $SU(2)_R$ gauge symmetry at high energies. We have 
considered both type-II and type-I seesaw scenarios. For simplicity, we have assumed in most 
scenarios that only the lightest sterile-neutrino mass eigenstate $N_4$ is kinematically available. 

For sterile-neutrino production and decay in DV searches, the essential benchmark parameters are the mass of the right-handed gauge boson $M_{W_R}$ and the ratio of vevs $\xi=\kappa^\prime/\kappa$, which sets the $W_L$-$W_R$ mixing strength.  
In scenarios with a relatively large $\xi=0.3$, hadronic processes may be significantly enhanced or suppressed through interference terms linear in $\xi$, depending on the parity configuration of initial and final-state particles.

In type-I seesaw scenarios, the left- and right-handed neutrino-mixing matrices $U_{\rm PMNS}$ and $U_R$ are, in principle, uncorrelated. 
To reduce the number of free parameters, we assumed these two matrices were equal, as in type-II seesaw scenarios. 
In this limit,  we note that for relatively light ($\lesssim 12$\,TeV) RH gauge-boson masses, minimal-mixing contributions induced by the non-zero Dirac mass term $M_D$ are always sub-leading compared to RH-current contributions so that the type-I and type-II scenarios have a very similar phenomenology.

By performing Monte-Carlo simulations, we have investigated the sensitivity reaches of the following (future) DV experiments: \texttt{ANUBIS}, \texttt{CODEX-b}, \texttt{DUNE}, \texttt{FACET}, \texttt{FASER(2)}, \texttt{MATHUSLA}, \texttt{MoEDAL-MAPP1(2)}, and \texttt{SHiP}. 
We have considered $B$-meson, $D$-meson, and kaon decays, where we have probed scenarios with and without the $W_L$-$W_R$ mixing. 
We highlight the sensitivity reach of the \texttt{DUNE} and \texttt{SHiP} experiments, which could probe sterile-neutrino signal events for RH gauge-boson masses $\sim 20$\,TeV.

The presence of relatively light Majorana neutrinos with non-standard interactions with SM fields can 
lead to large neutrinoless double beta decay rates. We have computed these rates using modern EFT methods. 
In both type-I and type-II scenarios, there are several distinct contributions to 
the $0\nu\beta\beta$ rates. In particular, an exciting interplay exists in the type-I scenario between 
contributions from light active neutrinos and the HNLs with RH interactions. In scenarios with small 
$W_L$-$W_R$ mixing, we find that the future DV searches at \texttt{DUNE} and \texttt{SHiP} will probe 
parameters space beyond the existing $0\nu\beta\beta$ bounds, including the future ton-scale program. 
Non-zero $W_L$-$W_R$ mixing tends to increase $0\nu\beta\beta$ rates while suppressing the DV-search 
sensitivity, making $0\nu\beta\beta$ searches more competitive.

The large separation of scales between the masses of the light sterile neutrinos and the heavy RH gauge bosons present in the mLRSM allows the description of the physics studied here from another perspective, the EFT approach.
The relevant EFT Lagrangians are $\nu$SMEFT and $\nu$LEFT, working above and below the EW scale, respectively.
We have performed the matching between the mLRSM and the EFT Lagrangians.
Previous studies applying EFT techniques for long-lived sterile neutrinos assume only specific Wilson coefficients with certain lepton- or quark-flavor configurations to be non-vanishing. 
While efficient, this approach considers scenarios that can be over-simplified from the view of UV completions.
We have chosen to focus on the well-motivated UV-complete mLRSM analysis, which provides a more comprehensive phenomenology and, as shown, probes BSM scales comparable to those predicted with the EFT approach.
However, an essential distinction between the two approaches is that a UV-complete model offers better predictability and allows for consistent consideration of complementary probes from sterile-neutrino-induced low-energy phenomena such as $0\nu\beta\beta$.

The main conclusion of this work is that in the mLRSM, the \texttt{LHC} far detectors, the \texttt{DUNE} near detector, and the beam-dump experiment \texttt{SHiP} at CERN can probe energy scales of the RH-current physics comparable to those that current and future \ovbb searches are sensitive to.
Both DV and \ovbb searches are, therefore, essential for scrutinizing the existence of long-lived GeV-scale sterile neutrinos.
Possible follow-up ideas of this work include investigating the correlation with constraints that can be set from Big Bang Nucleosynthesis. 
Additionally, it could be interesting to include sub-leading partonic processes in the DV searches and study their corresponding sensitivity results; in particular, they could have a more extensive mass reach on the sterile neutrino than the meson-decay events considered here.